\documentclass{aa}
\usepackage{txfonts}

\usepackage[dvipdfm]{graphicx}
\usepackage{graphicx}
\usepackage{color}
\graphicspath{{./}{fig1/}{fig2/}}

\newcommand{\sect}[1]{Sect.\,\ref{#1}}
\newcommand{\sects}[1]{Sects.\,\ref{#1}}
\newcommand{\fig}[1]{Fig.\,\ref{#1}}
\newcommand{\figs}[1]{Figs.\,\ref{#1}}

\usepackage{natbib,twoopt}

\def\sci{Science}
\title{Parametrization of coronal heating:
       spatial distribution and observable consequences}
\titlerunning{parametrization of coronal heating}
\author{T.~van~Wettum \inst{1,2}\and S.~Bingert\inst{1}\and H.~Peter\inst{1}}
\authorrunning{van Wettum, Bingert, \& Peter}
\institute{Max-Planck-Institut f{\"u}r Sonnensystemforschung, 
           37191 Katlenburg-Lindau, Germany,
           \hbox{vanwettum@mps.mpg.de} 
           \and
           Institut f{\"u}r Astrophysik, Universit{\"a}t G{\"o}ttingen,
           Friedrich-Hund-Platz 1,
           37077 G{\"o}ttingen, Germany}
\date{Received ... / Accepted ...}

  \abstract
  {}
  {We investigate the difference in the spatial distribution of the energy input for  parametrizations of different mechanisms to heat the corona of the Sun and possible impacts on the coronal emission.}
  {We use a 3D MHD model of a solar active region as a reference and compare the Ohmic-type heating in this model to parametrizations for alternating current (AC) and direct current (DC) heating models, in particular, we use Alfv\'en wave and MHD turbulence heating. We extract the quantities needed for these two parametrizations from the reference model and investigate the spatial distribution of the heat input in all three cases, globally and along individual field lines. To study differences in the resulting coronal emission we employ 1D loop models with a prescribed heat input based on the heating rate we extracted along a bundle of field lines.}  
  {On average, all heating implementations show a roughly drop of the heating rate with height. This also holds for individual field lines. While all mechanism show a concentration of the energy input towards the low parts of the atmosphere, for individual field lines the concentration towards the footpoints is much stronger for the DC mechanisms than for the Alfv\'en wave AC case. In contrast, the AC model gives a stronger concentration of the emission towards the footpoints. This is because the more homogeneous distribution of the energy input leads to higher coronal temperatures and a more extended transition region.}
  {The significant difference in the concentration of the heat input towards the foot points for the AC and DC mechanisms, and the pointed difference in the spatial distribution of the coronal emission for these cases shows that the two mechanisms should be discriminable by observations. Before drawing final conclusions, these parametrizations should be implemented in new 3D models in a more self-consistent way.}

\begin{document}

\maketitle

\section{Introduction}

The question of heating the corona to temperatures well in excess of the photosphere  is not yet fully answered, despite several decades of research. Since long, two categories based on magnetic mechanisms have been favoured, alternating current (AC) and direct current (DC) heating. In the former case the magnetic field is driven faster than the Alfv\'en crossing time, giving rise to magneto-acoustic waves. In the latter the driving is more gradual and leads to a braiding of field lines. In both cases in the end the induced (alternating or direct) currents are dissipated and thus heat the coronal plasma. There is ample observational evidence for the presence of Alfv\'enic waves in the corona that could lead to AC heating \citep[e.g.][]{Tomczyk+al:2007}. Recently there have been  claims that braiding of field-lines has been observed \citep{Cirtain+al:2013}, which would lead to DC heating. This would confirm the original theoretical braiding concept by \cite{Parker72} or the flux-tube tectonics model \citep{Priest+al:2002}. However, the actual \emph{microscopic} dissipation process is still not answered by any of these concepts.

According to classical transport theory, the actual dissipation should happen on very small scales of a meter and below, which is because of the low value of the magnetic diffusivity \citep[e.g.][]{Boyd+Sanderson:2003}.
It is clear that such small scales cannot be reached in a model that is built to describe actually observed solar structures, e.g. an active region,  that encompasses about 100\,Mm in each spatial dimension. Accounting for the spatial complexity of the corona one needs to employ a three-dimensional model, which then limits the possible spatial resolution in the models the order of 10\,km to 100\,km (depending on the size of the computational domain). This is a bit better than what can be achieved with current observations, where the spatial resolution in the corona is mostly limited to about 1\,Mm. Therefore, to understand the appearance of the corona on the observable scales one has to employ more or less well founded parametrizations for the spatio-temporal distribution of the heat input. If different parametrization, e.g.\ for AC and DC heating, result in coronae that look different, a comparison to observations can be employed to test which of the parametrizations does fit better. This is then an (indirect) confirmation which heating mechanism is dominant in the solar structure under investigation.

In this investigation we will use a 3D MHD model of the corona to study the spatial distribution of the energy input to heat the corona for different parametrizations. In particular we will apply the AC mechanism of energy input through Alfv\'en waves \citep{balleg2011} and the DC mechanism of MHD turbulence \citep{Rappazzo+al:2008}. As a reference, we will use a model that is based on the concept of coronal heating through field-line
braiding, as suggested by \cite{Parker72}, where magnetic field lines are braided through the granular motion
in the photosphere. The currents induced by this process are then dissipated through Ohmic dissipation. The feasibility to maintain a hot loop-dominated
corona through this mechanism, has been shown by \cite{Gudiksen02,Gudiksen05b,Gudiksen05a}. Since then it has been shown that the transition region and coronal emission synthesized from such models matches well to actual observations in a statistical sense \citep{Peter+al:2004,Peter+al:2006}. In particular this type of models provides a good understanding of the transition region red-shifts  \citep{Zacharias2011} and the coronal blue-shifts \citep{Hansteen+al:2010}, even though the interpretation of these models differs. Furthermore these models provide detailed information on the distribution of the heat input in space, time and energy \citep{Bingert+Peter:2011,Bingert+Peter:2013}, produce cool structures that are ejected into the corona \citep{Zacharias+al:2011.blob} and give an explanation for the constant cross-section of extreme UV loops \citep{Peter+Bingert:2012}. Therefore we can consider these type of coronal models realistic in the sense that they reproduce major  features we see in actual observations.
The main goal of this study will be to investigate to what extent the results depend on the assumed form of the Ohmic heating, or if other (parametrized) types of heating would produce similar results.\\

We will first give an introduction of the 3D MHD model we use and how we parametrize the different mechanisms for the heating in \sect{S:models}. We then discuss the general spatial distribution of the heating rate averaged and along individual field lines (\sect{S:spat.distr}) before we turn to a comparison of the different heating mechanisms (\sect{S:comp.heat}). Finally, we present some simple 1D loop model to investigate the observational consequences for the different mechanism in \sect{S:one-d}.

\section{Models of atmospheric structure and analysis}\label{S:models}

\subsection{Full 3D MHD with Ohmic heating}
\label{sect:3D_MHD}

In this investigation we employ the 3D MHD simulation \cite{Bingert+Peter:2011}. 
This numerical experiment uses Pencil Code \citep{brandenburg02} to solve the time-dependent MHD equations in three dimensions. The computational domain encompasses the volume from the photosphere to the corona of a small active-region. The computational domain stretches $50 \times 50$\,Mm$^2$ in the horizontal direction and 30\,Mm in the vertical direction covered by a 128$^3$ grid. In the horizontal direction the box is periodic. At the bottom boundary the evolution of the magnetic field in the simulation is driven through a horizontal driver, which has the same properties as the horizontal granular motions (size, velocities, lifetime, power spectrum). The model solves the full energy equation which includes, in addition to Ohmic heating, the optical thin radiative losses \citep{Cook89}, and in particular the Spitzer heat conduction parallel to the magnetic field \citep{Spitzer62}. The latter is pivotal to get the proper pressure of the corona set self-consistently. The full set-up has been discussed in detail by \cite{Bingert+Peter:2011}.

The Ohmic heating rate (per volume, e.g. measured in W/m$^3$) through the dissipation of currents that are induced by the footpoint motions is given by  

\begin{equation}
Q_{\rm Ohm} = \eta \mu_0 \mathbf{j}^2 ,
\label{eq:Ohm}
\end{equation}
where $\mu_0$ is the magnetic permeability in vacuum and the current $\mathbf{j}$ is given by the curl of the magnetic field, $\nabla{\times}\mathbf{B}/\mu_0$. 
The value of the magnetic resistivity is set to, $\eta=10^{10}\ \mathrm{m}^2/\mathrm{s}$. This choice is made to ensure that the magnetic Reynolds number (when using the grid spacing as a length scale) is of order unity, as discussed in some detail by \cite{Bingert+Peter:2011}. By this the currents will be dissipated at the grid spacing. 

After some initialization the model corona reaches a quasi-stationary state,
i.e. the corona continues to change its thermal structure and harbours strong flows, but in a statistical sense (when horizontally averaging over the box) the quantities remain comparably constant. In this state
 the radiative losses and the heat conduction balance the heat input by Ohmic dissipation and, through to a lesser extend, viscous heating. Therefore the \emph{average} vertical profiles of temperature and pressure of the corona are mainly determined by the energy inserted into the corona through the Poynting flux. The magnetic diffusivity determines how efficient and at which scales the energy is deposited. For the purpose of this study we therefore consider $\eta \mu_0 \mathbf{j}^2$ as a \emph{parametrization} of Ohmic heating.


\subsection{Reduced MHD models for other heating mechanisms}

We compare this parametrization of Ohmic heating with two parametrizations  of coronal heating suggested recently. Both of these suggestions are based on work done with reduced MHD models, where a high resolution can be reached by reducing the equations to only consider the deviations from a (steady) background.
Both models solve the heating along a coronal loop and are driven by photospheric motions at the boundaries. As a trade-off for the high resolutions reached by these simulations, concessions are made in regard to certain physical aspects as gravity, conduction, or even the presence of a temperature in the energy equation.
 Therefore they do not provide a possibility to compare the results directly with observables, i.e., through synthesized coronal emission or Doppler shifts. For  this one would have to include in particular the heat conduction to  allow the coronal pressure to self-consistently adjust to the energy input. 

These two heating parametrizations further investigated in this paper are based on heating through Alfv\'en wave dissipation and MHD turbulence.

\emph{Alfv\'en waves:} This  mechanism operates through dissipation of Alfv\'en waves high in the corona, which are excited in the photosphere. 
 In their model \cite{balleg2011} encompass a gravitationally stratified flux tube, with a velocity driver at the bottom boundary that excites the Alfv\'en waves. These propagate upwards into the corona and and are dissipated there. The resulting loss of magnetic energy is then equated to the energy input into the corona. Based on their model results  \cite{balleg2011} derived the following parametrization (their Eq.\,63) of heating rate per volume,  
\begin{eqnarray}
Q_{\rm alf}\propto B^{0.55} L^{-0.92} v_{\rm rms}^{1.65} ~,
\label{eq:balleg}
\end{eqnarray}
where $v_\mathrm{rms}$ is the (root-mean-squared) velocity of the foot-point motions, $ L$ the length of the magnetic field line (or loop) and $ B$ the local magnetic field strength.
We will refer to this parametrization as ``Alfv\'enic heating''.

\emph{MHD turbulence:} This heating mechanism acts through an anisotropic turbulent cascade forming thin elongated current sheets. In their reduced MHD model \cite{Rappazzo+al:2008} analysed the energy input through this process and derived a parametrization (their Eq. 68). One of their exponents in the parametrization depends weakly on the Alfv\'en speed, $(\alpha{+}1)/(\alpha{+}2)$ ranges from 0.6 to 0.9 for Alfv\'en speeds from 50\,km/s to 1000\,km/s. For the following we thus adopted a value of 0.75. Then one can rewrite the parametrization of  \cite{Rappazzo+al:2008} as
\begin{equation}
Q_{\rm turb}\propto B^{1.75} L^{-1.75} \rho^{0.125} v_{\rm rms}^{1.25} \ell^{0.75}  ~,
\label{eq:rappazzo}
\end{equation}
where $\rho$ is the local density of the plasma and the other parameters are as above. As also suggested by \cite{Rappazzo+al:2008}  we assume the injection length $\ell$, i.e.\   the scale of granules, to be constant. We will refer to this as ``turbulent heating''.

\subsection{Implementation of the heating parametrizations}\label{S:implementation}

We will use these parametrizations, Eqs. (\ref{eq:Ohm}) to (\ref{eq:rappazzo}), in order to investigate the spatial distribution of heat input into magnetic loops in the corona. For this we will extract  the relevant parameters from the 3D MHD model along magnetic field lines.


The local magnetic field strength $B$, the mass density $\rho$,  and  the currents $\mathbf{j}{=}\nabla{\times}\mathbf{B}/\mu_0$ can be easily calculated at each grid point of the 3D MHD model. To determine the length $L$ of the field line passing through each grid point and the rms-velocity $v_{\rm{rms}}$ at the foot points of these field lines, we trace the field line crossing each grid point in the computational domain. We consider only closed field lines, i.e. those for which the tracing from the grid point in both directions ends up at the bottom boundary. For the parametrization we actually use the average of  $v_{\rm{rms}}$ at
both foot points.
After the calculation of the trajectories of the field lines we interpolate $B$, $\rho$ and $\mathbf{j}^2$ along each of the field lines. Through this we can investigate the distribution of the heat input for the parametrizations (\ref{eq:Ohm}) to (\ref{eq:rappazzo}) along individual field lines without running a new simulation with a different heating term in the energy equation (see \sect{S:field.lines}).

The case of the Ohmic dissipation (\ref{eq:Ohm}) will be self-consistent, because the 3D MHD model we use to extract $Q_{\rm{Ohm}}$  uses exactly this form of the heating.
Consequently,
the other two cases will not be fully self-consistent. However, for the Alfv\'enic heating $Q_{\rm{alf}}$  (\ref{eq:balleg}) depending only on magnetic quantities (field strength and field line length), this should not be a significant problem, at least not in the coronal part of the computational domain, which is what we are interested in. Here the plasma-$\beta$ is much smaller than unity and thus the magnetic field dominates the plasma. Consequently, the heat input will change only the thermal structure, but not the magnetic structure (or only to a small extent)  --- there is no significant back-coupling from the heating rate to the magnetic structure. The situation is less favourable for the turbulent heating $Q_{\rm{turb}}$ (\ref{eq:rappazzo}), because this depends also on the thermal structure through the mass density. However the dependence ($\rho^{0.125}$) is only weak, which is why this inconsistency should be mostly acceptable. We will come back to this issue in \sect{S:random.set}. A set-up with a fully self-consistent treatment of the different heating parametrizations in the framework of new 3D MHD\ models is planned for future work.


\section{Spatial distribution of the heat input}\label{S:spat.distr}

\subsection{Horizontal averages}\label{S:hor.avg}

\begin{figure}
        \includegraphics[width=\columnwidth]{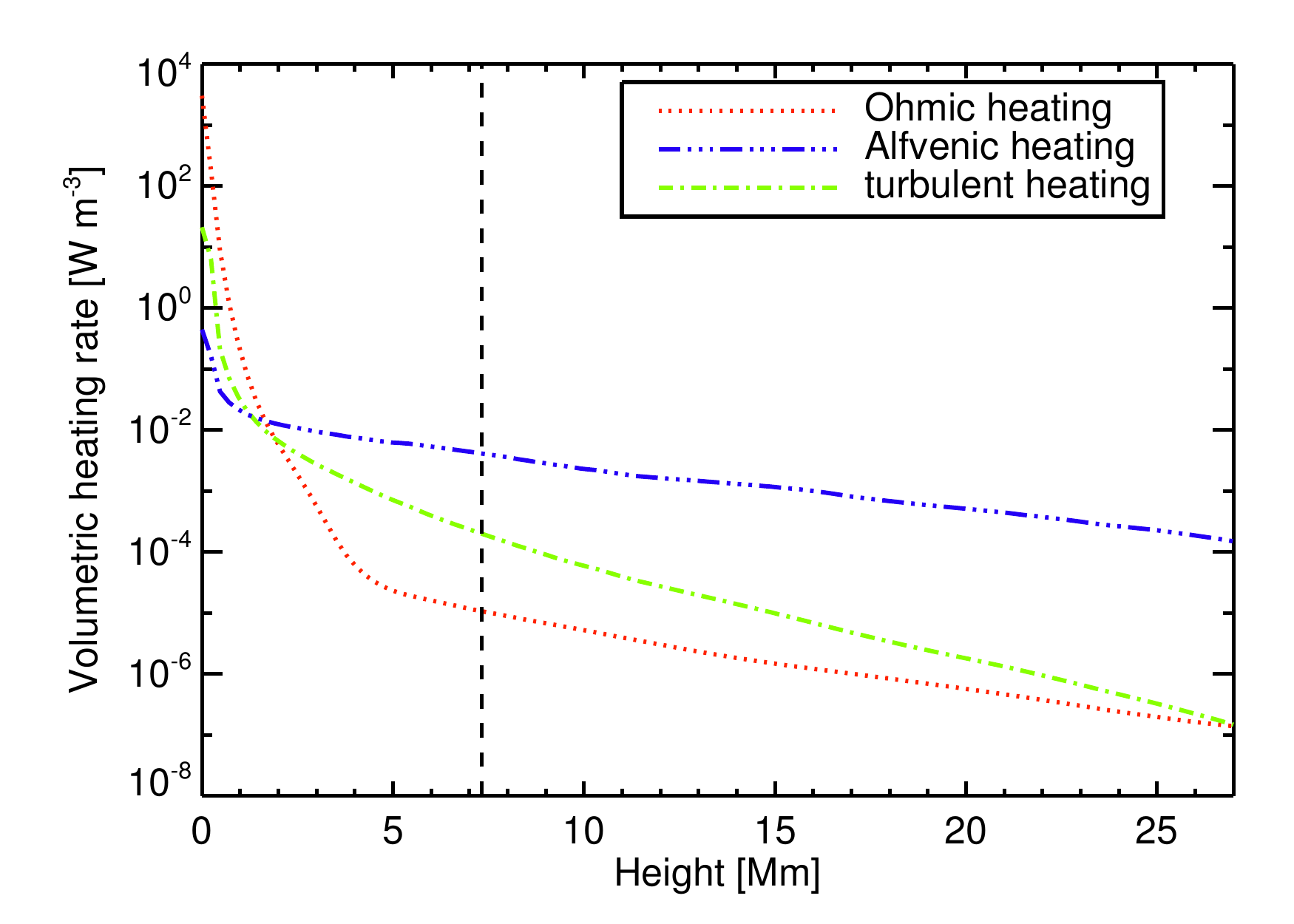}
\caption{Horizontally averaged heating rates for different parametrizations.
The dashed line indicates the average height of the base of the corona ($\log{T
[K]}{=}5.5$). The Alfv\'enic and turbulent heating rates are in arbitrary
units.
See \sect{S:hor.avg}.
\label{fig:av_heating}}
\end{figure}

Using the method described above, the heating rate for all three parametrizations  Eqs. (\ref{eq:Ohm}) to (\ref{eq:rappazzo}) at every grid point in the simulation box is calculated. Subsequently we determine the horizontally averaged heating rates (as a function of height) shown in Fig. \ref{fig:av_heating}. 
Only for Ohmic heating we use physical units, the turbulent and Alfv\'enic heating rates are plotted in arbitrary units.

%
%

All three parametrizations share the 
property that they drop roughly exponentially in the coronal part of the volume. For the Ohmic heating case this is well known from previous studies. That this exponential drop is a common feature for all these three processes underlines the result (in part based on observations) that the energy input into the corona should be concentrated towards the footpoints \citep[e.g.][]{Aschwanden+al:2007}. While the (exponential) scale height for the drop of the heating rate is about 5\,Mm for the Ohmic and Alfv\'enic heating, it is only 3\,Mm for the turbulent heating. Thus the turbulent heating drops slightly faster than the two other mechanisms. 

This common exponential drop for the three mechanisms is of interest, because different (1D loop) models made different assumption for the spatial distribution. 
While many models have assumed this exponentially dropping heat input \citep[e.g.][]{Serio81,Mueller+al:2003} there are also numerous models that assume a spatially constant heating rate \citep[e.g.][]{Patsourakos06,Klimchuk06}.

Based on the horizontal averages alone as shown in \fig{fig:av_heating} one cannot really conclude that the distribution of the heat input along each magnetic field line is non-constant, but dropping with height.
Since the heating rates  depend inversely on the loop length, it could be the result of stronger heating along the short field lines in the lower regions, even if the heating rate along each individual field line is constant. To determine if the drop of the horizontally averaged heating rate is because the heating drops with height for each field line or if it is because longer field lines are heated less, we will investigate individual field lines.

\subsection{Variation along individual field-lines}\label{S:field.lines}

%
\begin{figure}
\includegraphics[width=\columnwidth]{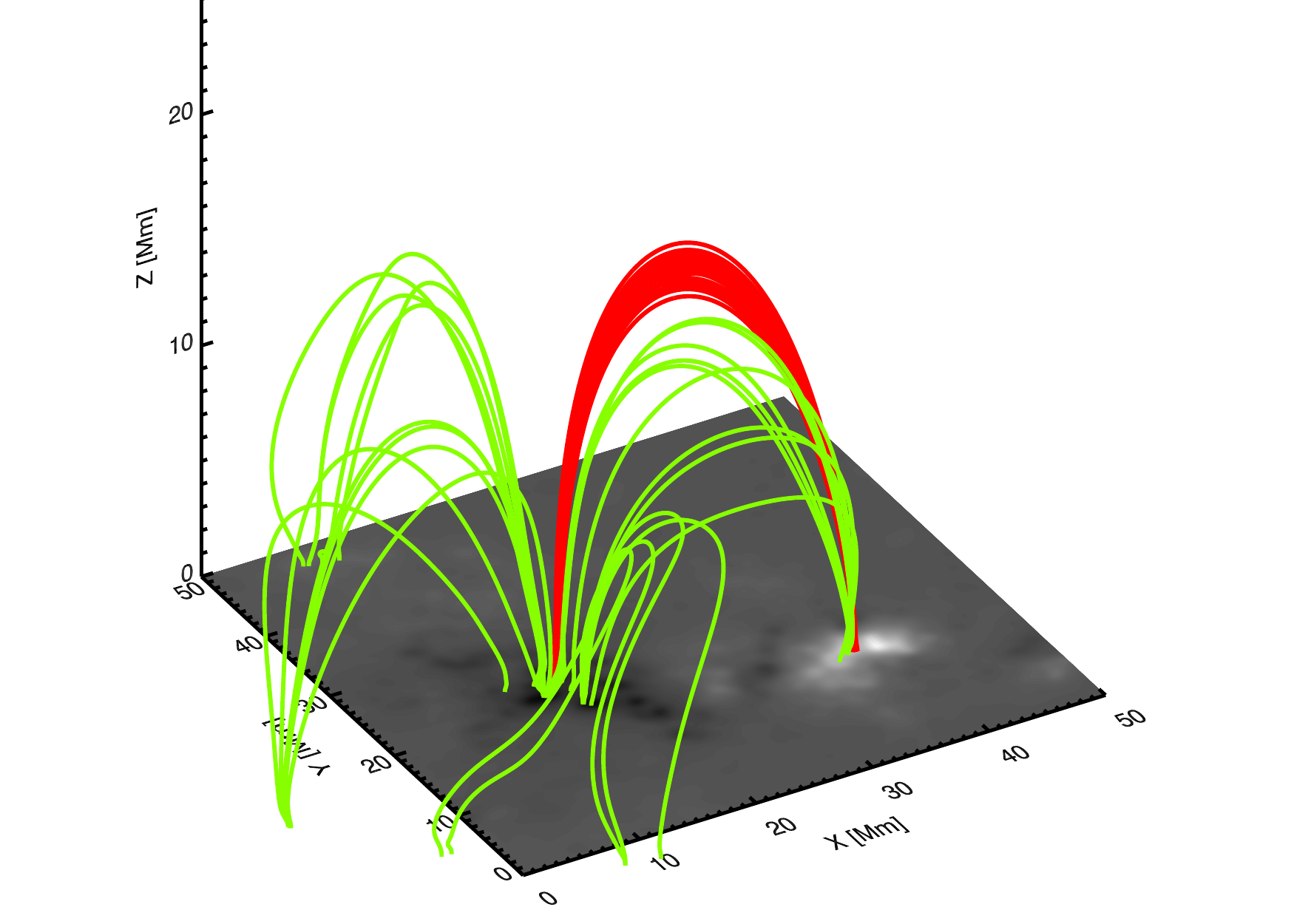}
\caption{Three-dimensional illustration of selected field lines for this
investigation. The red lines belong to the ``bright loop'' selection, and
the green ones to  selection of the ``random set'' of field lines. The gray-scale image
 shows the magnetogram at the bottom boundary of the simulation (that is
based on observations). \label{fig:loops-mid}}
\end{figure}
%

We make two selections of field lines based on their maximum temperature and density along the loop. Other selections were investigated, but the principle results did not significantly differ from these two selections and thus we will limit the discussion to only these two.
The first selection of field lines is roughly co-spatial with a bright loop in synthesized coronal emission that was studied in detail by \cite{Peter+Bingert:2012}. These field lines have lengths between 45 and 50 Mm, a maximum temperature exceeding $\log T\,\rm{[K]}{=}$\,6.15 and a minimum density larger than $10^{8.5}$ particles per cm$^3$. We will refer to this set as the ``bright loop''. Figure \ref{fig:loops-mid} shows an illustration of the computational domain with the bottom boundary magnetic field indicated. The ``bright loop'' is plotted here as the dense cluster of red lines.\\
The second selection includes field lines which are not limited to the central bright loop, but are more randomly distributed throughout the corona. The requirements for the density are similar to those of the bright loop, but a lower maximum temperature, $\log T/\rm{[K]}=$\,6.05. The lengths of the field lines of this selection range from 40 to 45 Mm. This more random selection of field lines will provide a test to confirm whether the heating profiles calculated along the bright loop set are typical heating rates. We will refer to this set as the ``random set''. It is drawn in Fig. \ref{fig:loops-mid} as the green lines. From both selections a sub-set of 25 randomly selected lines have been used for the plots in this paper to not clutter the panels.

\begin{figure} 
\centerline{\includegraphics[width=0.8\columnwidth]{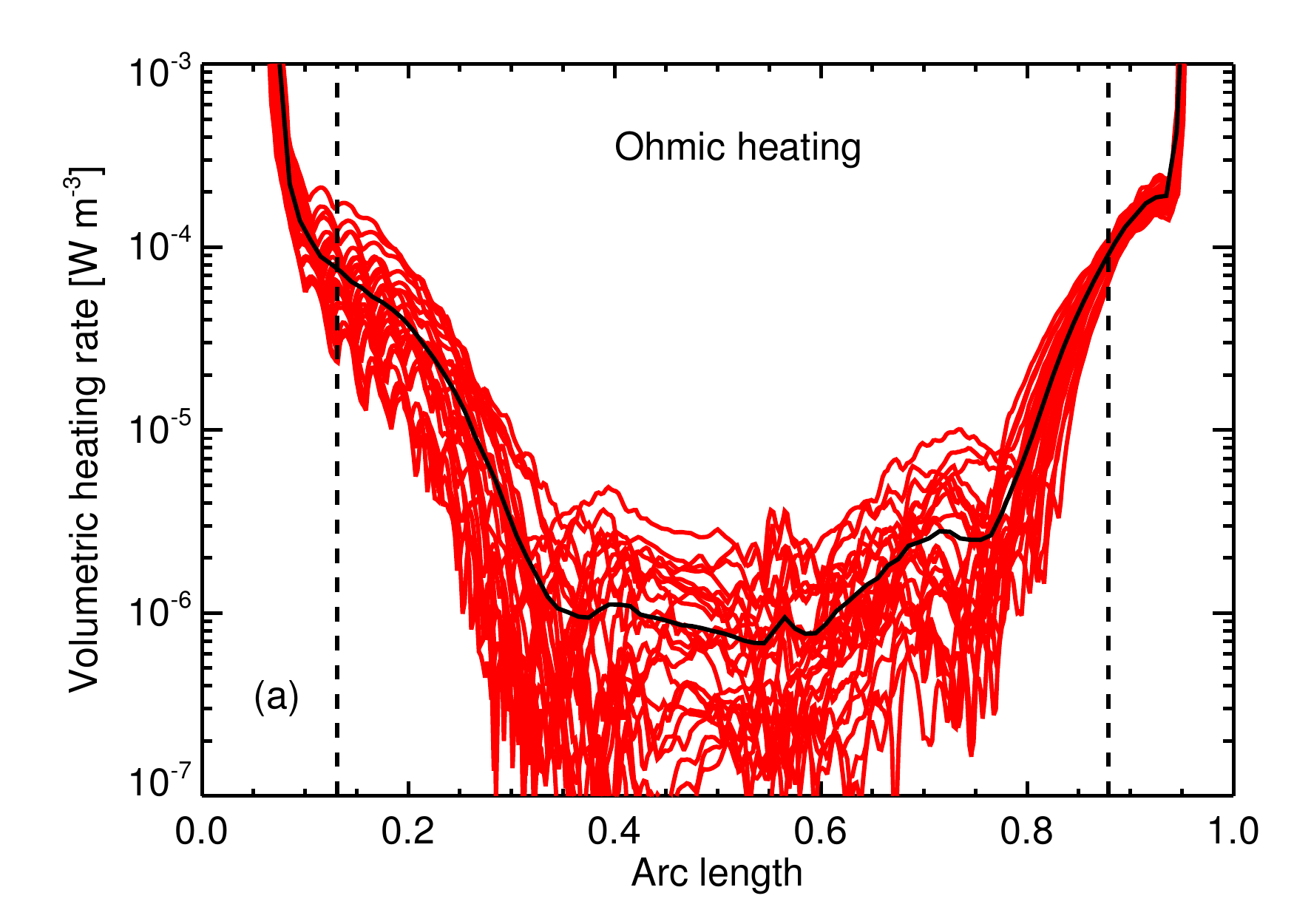}}
\centerline{\includegraphics[width=0.8\columnwidth]{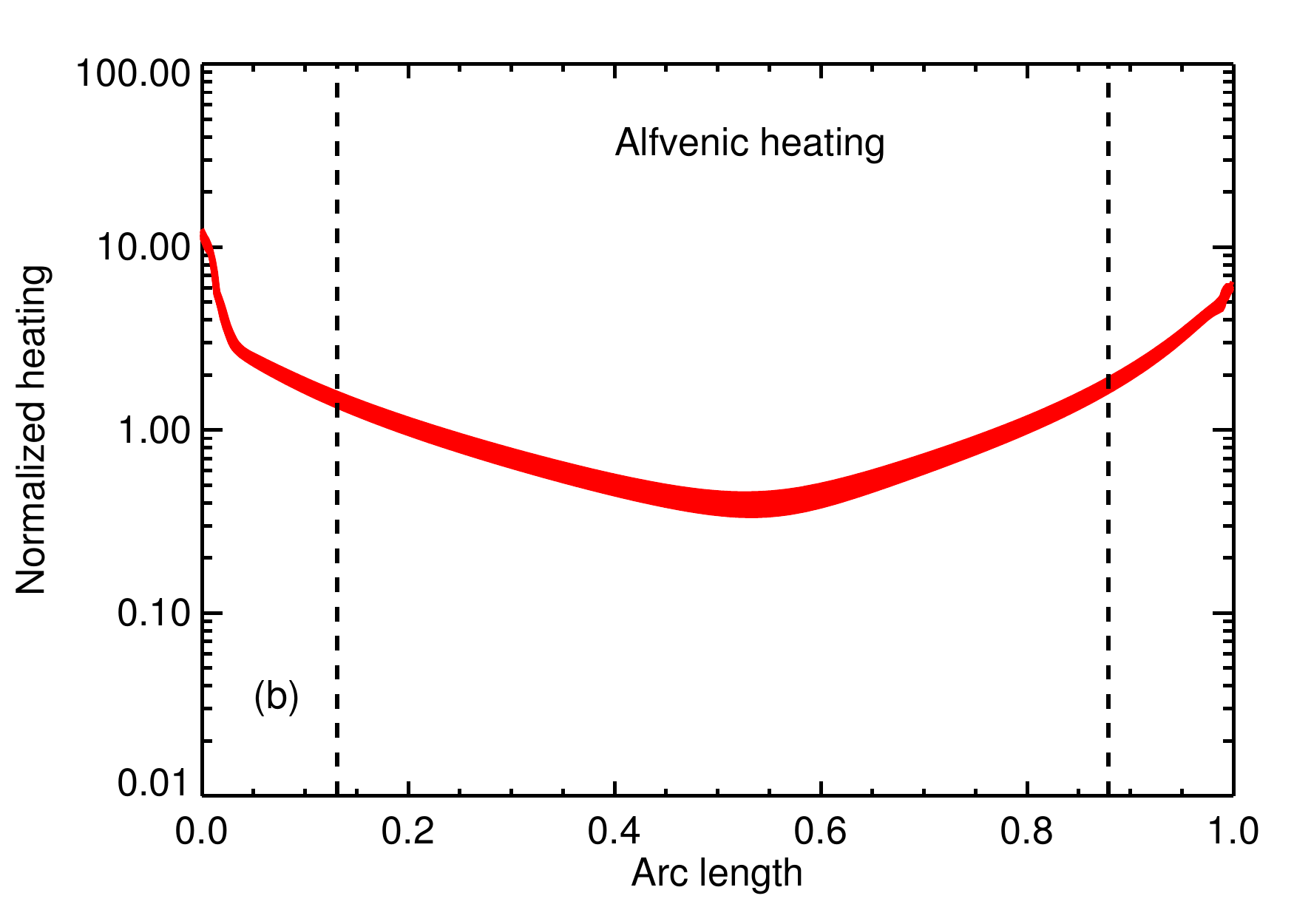}}
\centerline{\includegraphics[width=0.8\columnwidth]{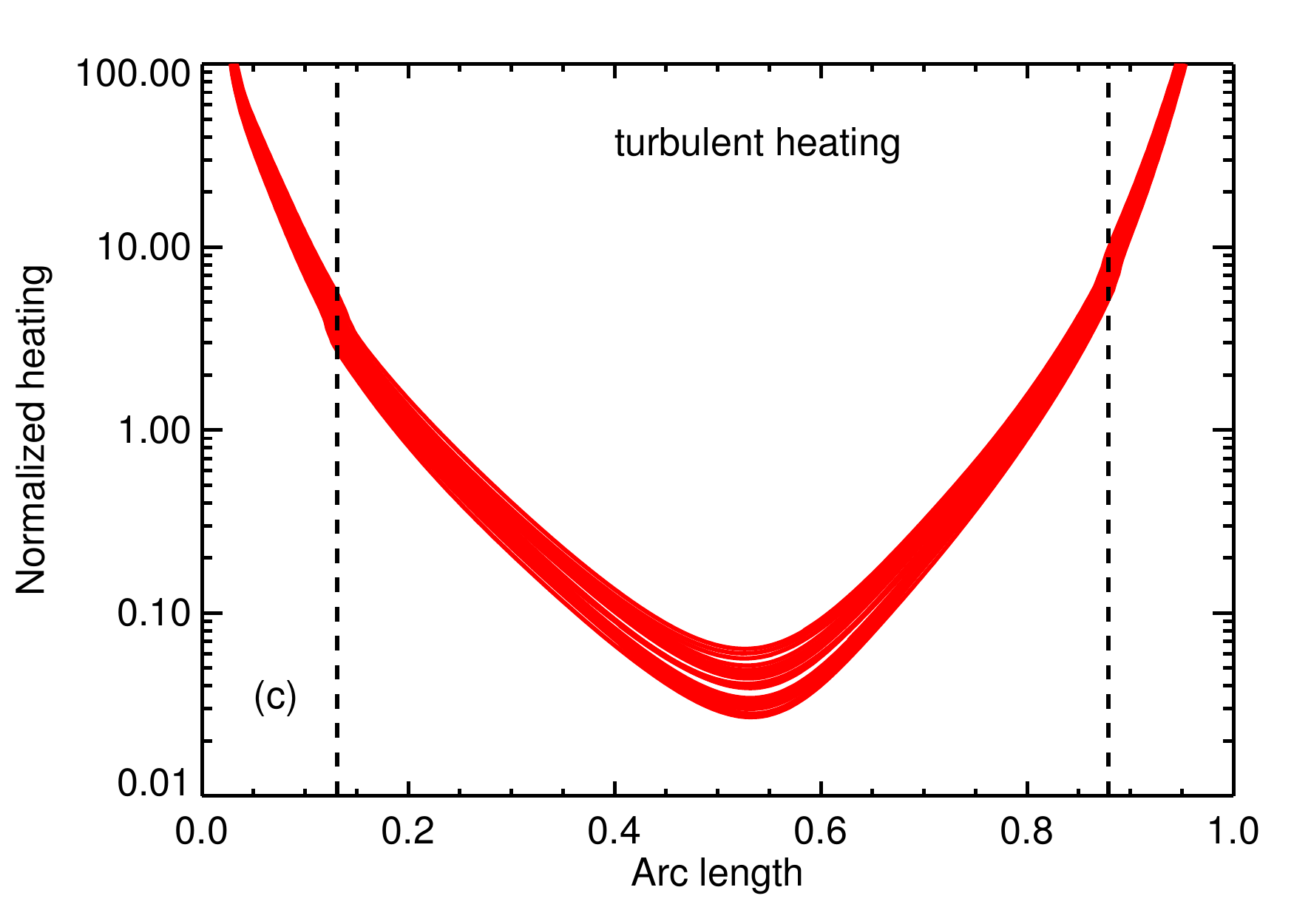}}
\caption{Volumetric heating rates along individual field lines of the ``bright loop''
set, marked in red in Fig. \ref{fig:loops-mid}. The lengths of the field lines are normalized to unity. The black thick line
in the top panel shows the average of the Ohmic heating for  the selected
field lines. The Ohmic heating rates are averaged over 5 minutes to reduce
the effect of transient events. The dashed lines indicate the average position
of the  coronal base at  $\log{T}[\textrm{K}]{=}5.5$ for the selected field lines. The Alfv\'enic and the turbulent heating are plotted normalized to the heat input just below the coronal base. All panels cover the heating rate over four orders of magnitude on the ordinate.
See \sects{S:field.lines} and \ref {sect:hot_dense}.
\label{fig:set1_heating}}
\end{figure}

The results of the calculated heating rates along each of the the loops in these subsets are plotted in \figs{fig:set1_heating} and \ref{fig:set2_heating}. It is clear that the heating rate drops for each individual field line, irrespective of the heating mechanism we considered here (we discuss the differences between the mechanism in the following section). This shows that the \emph{horizontally averaged} heating rates shown in \fig{fig:av_heating} are \emph{not} an artefact of averaging over different loops with different length, while the heat input is constant along each loop. Also if we test different sets of loops (or for that matter, all loops reaching up into the corona) we find this result that the heat input drops strongly with height along each individual field line. This does not fully rule out that some loops might be heated constant in space, but at least for the mechanisms checked here we do not find this.

%
%
%
\section{Comparison of the heating mechanisms}\label{S:comp.heat}
\subsection{Magnetic field lines in a ``bright loop''} \label{sect:hot_dense}

For
the comparison of the parametrizations for Ohmic, Alfv\'enic, and turbulent heating we first investigate the set of field lines associated with a ``bright loop'' as defined in \sect{S:field.lines} and shown in red in   
 \fig{fig:loops-mid}. The volumetric heating rates for the three parametrizations along individual field lines in the bright loop are shown in \fig{fig:set1_heating}. We plot this as a function of the  arc length along the field line, where the length of each field line is normalized to unity. The field lines in the set differ in length by 10\% at most.

The most striking difference is that the Ohmic heating varies much stronger than the Alfv\'enic and turbulent heating on \emph{small} scales (smaller than a couple of \% of the field-line length). The reason for this is simply found in the fact that the Ohmic heating depends on the spatial derivatives of the magnetic field (actually, the square thereof). Naturally, these show much stronger small-scale (but well resolved) variations than the magnetic field itself. The original spatial variation of
the Alfv\'enic and turbulent heating rate in the respective numerical models \citep{balleg2011,Rappazzo+al:2008} also shows a stronger spatial variation. Only when considering the average behaviour to derive the parametrizations as a function of $B$ and other quantities the heating rate becomes smooth.

Apart from the small-scale variation of the (smoothed) Ohmic and turbulent heating along the magnetic field lines is rather similar. In both cases the heating rate drops from the base of the corona (indicated in \fig{fig:set1_heating} by vertical dashed lines) to the loop apex by about a factor of 150 to 200. This is not too surprising, because the \cite{Rappazzo+al:2008} 3D reduced MHD model for the turbulent heating is, in principle, quite similar to our 3D MHD model for the Ohmic heating \citep{Bingert+Peter:2011}. In both cases the foot points are smoothly driven at boundaries, which braids the magnetic field and induces currents. The reduced MHD model lacks the realistic set-up and the proper treatment of the energy equation to get the coronal pressure correct, but it can afford a much higher resolution in the numerical experiment and properly resolves the turbulent nature of the dissipation process. It is reassuring that these two models provide results for the heating rate that is not too different.

In contrast, the results for the Alfv\'enic heating following the \cite{balleg2011} parametrization shows a different drop of the heating rate.
From the coronal base to the top of the loop, the heating rate drops only by a factor of five to six (middle panel of \fig{fig:set1_heating}). Comparing the Alfv\'enic and turbulent parametrizations (\ref{eq:balleg}) and (\ref{eq:rappazzo})
makes clear that the magnetic field $B$ makes the difference. The lengths $L$ of the field lines in the set are the same within 10\%, the horizontal velocities at the footpoints $v_{\rm rms}$ due to the granulation cover only a small range and the drop of the density $\rho$ is not very important because of the comparable large barometric scale height (and the turbulent heating depends only weakly on $\rho$). However, the drop of the magnetic field from the coronal base to the loop apex by a factor of about 20 is mainly responsible to the large drop of the turbulent heating ($20^{1.75}{\approx}190$) and the only small drop of the Alfv\'enic heating ($20^{0.55}{\approx}5$).

\begin{figure} 
\centerline{\includegraphics[width=0.8 \columnwidth]{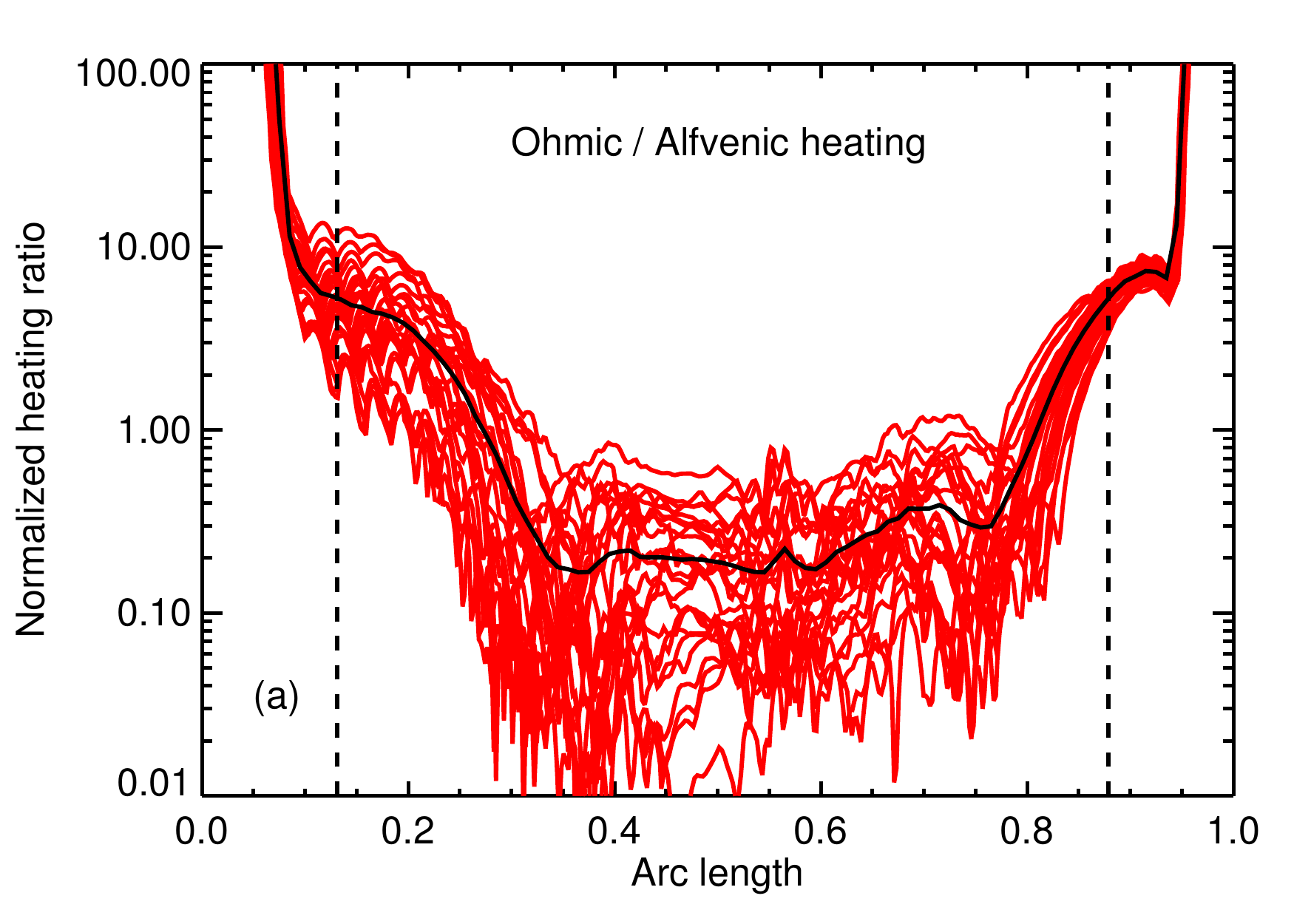}}
\centerline{\includegraphics[width=0.8 \columnwidth]{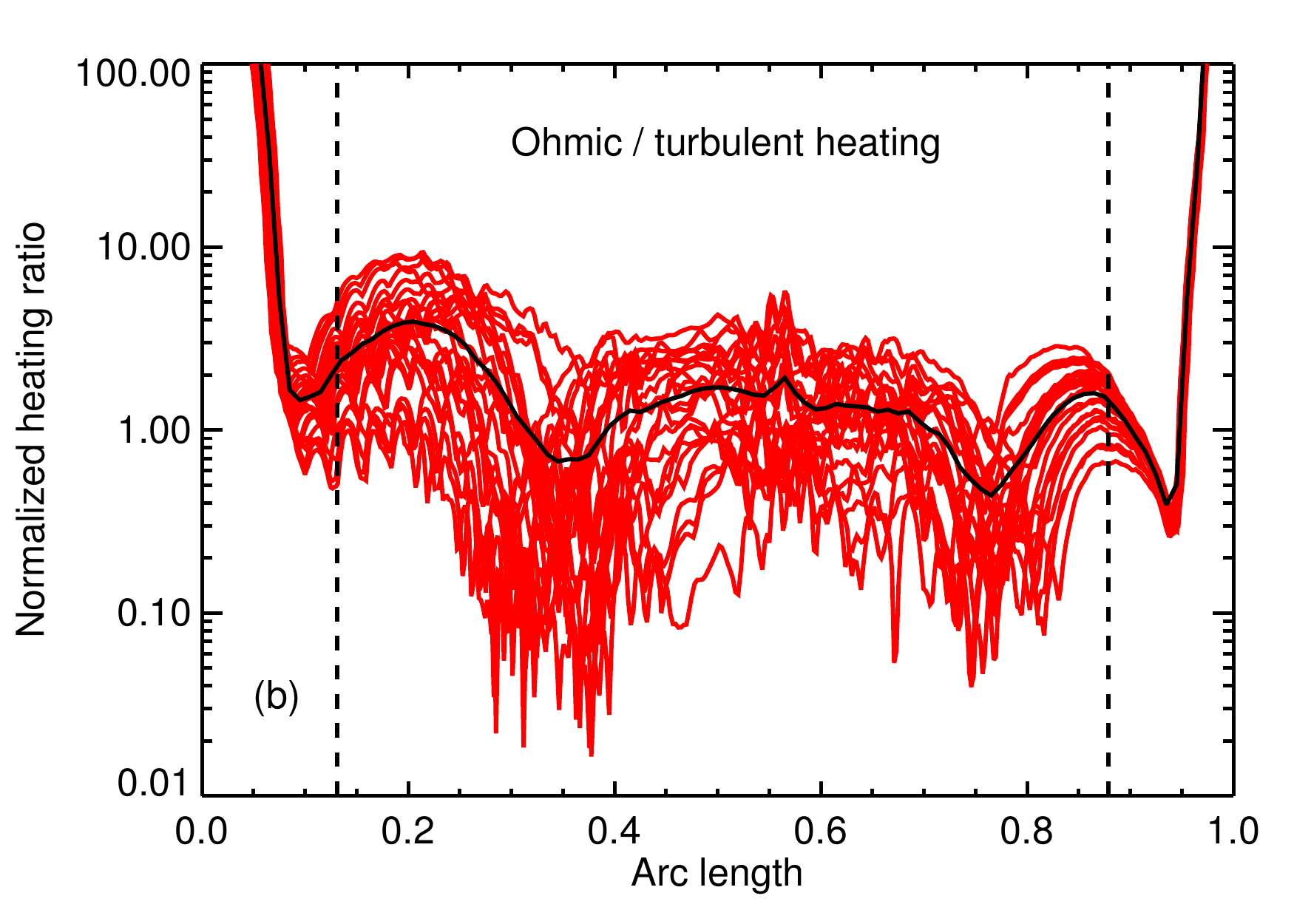}}
 \caption{Normalized ratio of  Ohmic heating to Alfv\'enic and turbulent heating (volumetric heating rates) for the field lines of the ``bright
loop''
set, marked in red in Fig. \ref{fig:loops-mid}. The lengths of the field
lines are normalized to unity. The black thick lines
show the average of the ratios for  the selected
field lines. The dashed lines indicate the average position
of the  coronal base at  $\log{T}[\textrm{K}]{=}5.5$ for the selected
field lines. Both panels show the ratio over a range of four orders of magnitude.
See \sect{sect:hot_dense}.
\label{fig:set1_fractions}}
\end{figure}

To highlight the differences in the spatial distribution of the three heating parametrizations, we plot  the ratio of the Ohmic to the Alfv\'enic and the  turbulent heating in Fig. \ref{fig:set1_fractions}. This underlines that (on average) the turbulent heating is quite similar to the Ohmic heating. They both show a much stronger concentration towards the footpoints, which is mainly because of the different dependence on the magnetic field strength. The heating rates also differ quite significantly below the base of the corona. This is not surprising, because the parametrizations for the Alfv\'enic and turbulent heating are derived for the corona. So taking them seriously in the chromosphere would be over-stretching these approximations. The 3D MHD model with the Ohmic heating shows a much stronger energy input in the chromosphere, which is because of the strong shearing of the magnetic field in the lower denser part of the atmosphere, where plasma-$\beta$ is no longer smaller than unity. 

%
%

\subsection{The ``random set'' of field lines}\label{S:random.set}

\begin{figure} 
\centerline{\includegraphics[width=0.8 \columnwidth]{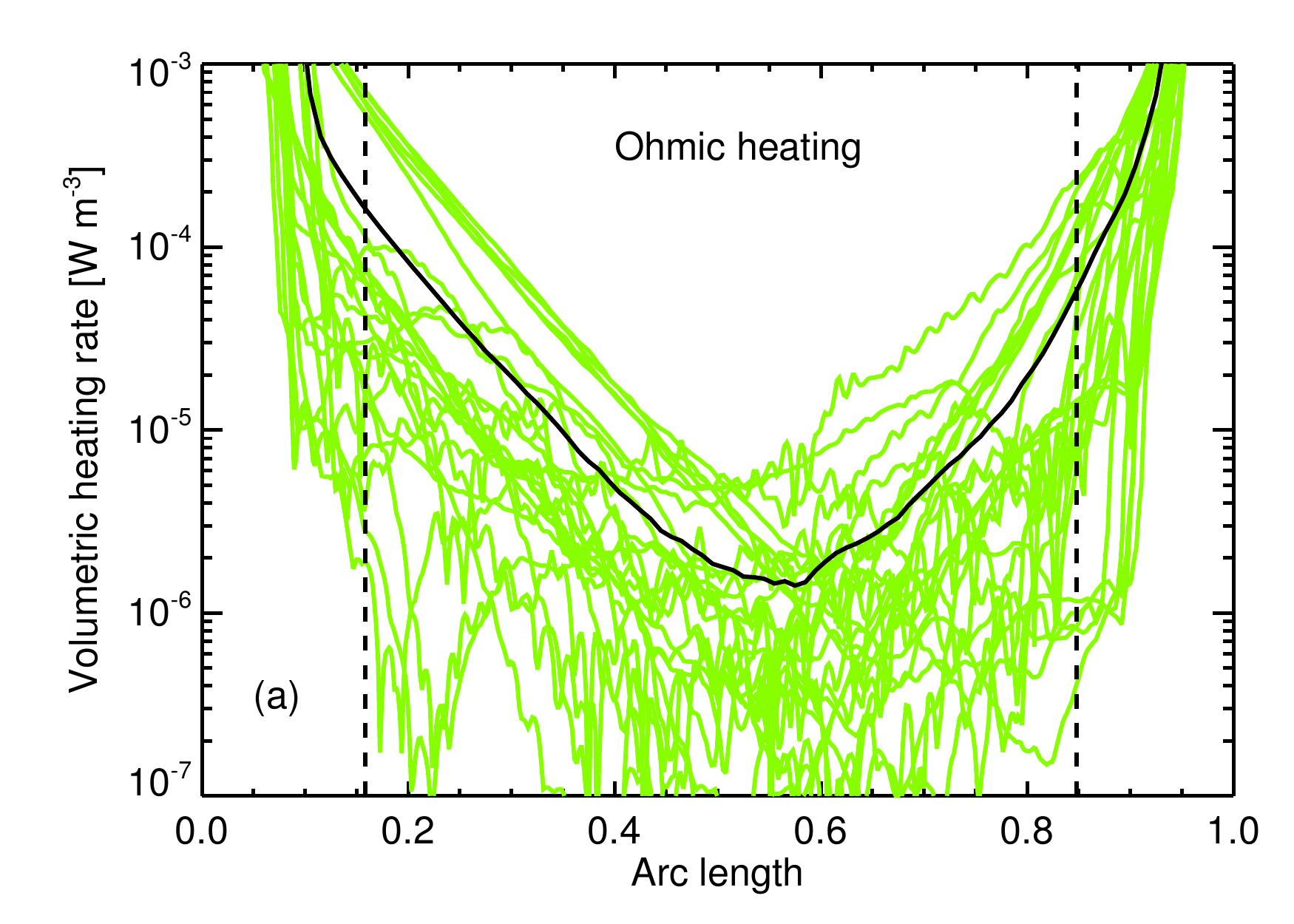}}
\centerline{\includegraphics[width=0.8 \columnwidth]{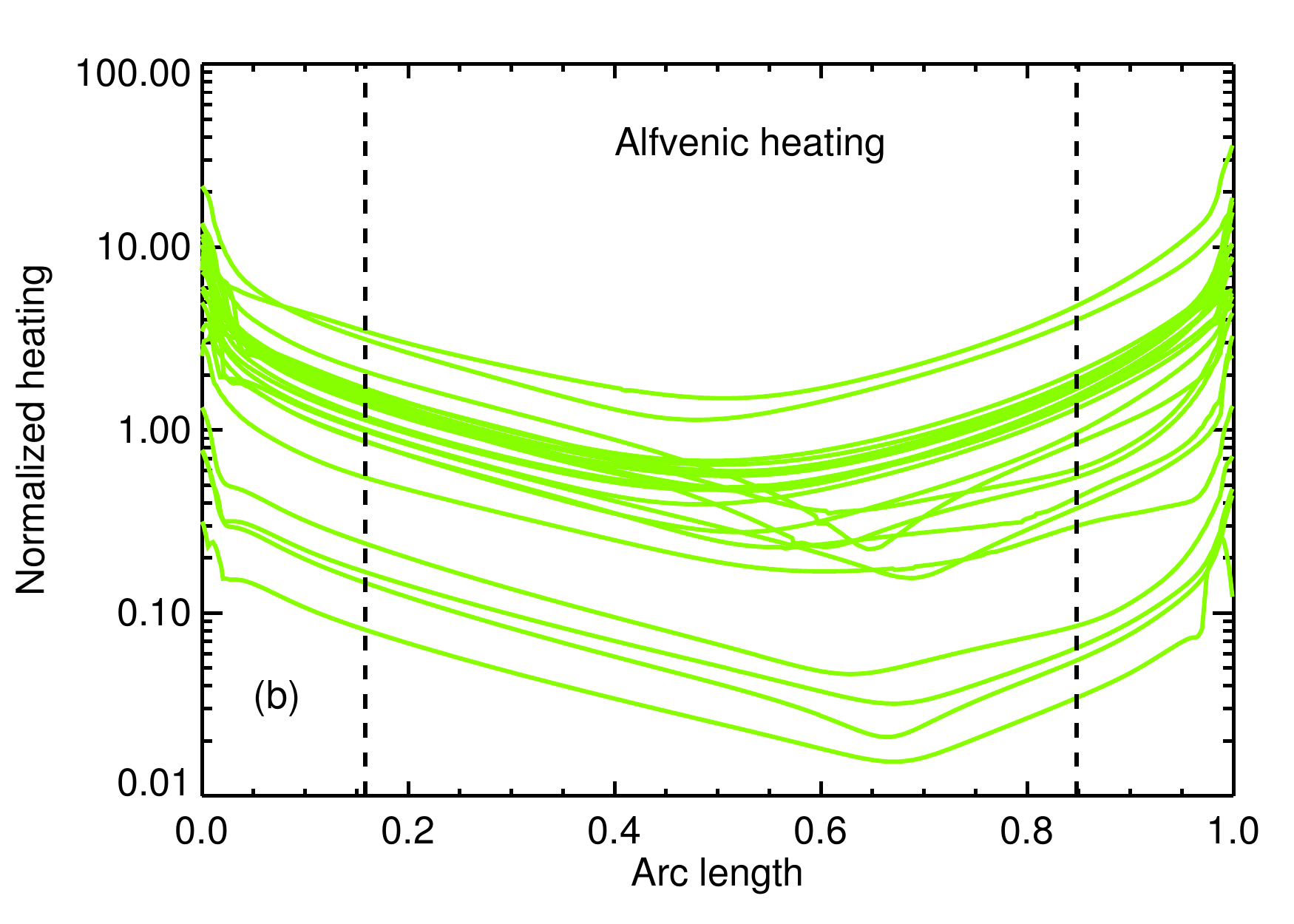}}
\centerline{\includegraphics[width=0.8 \columnwidth]{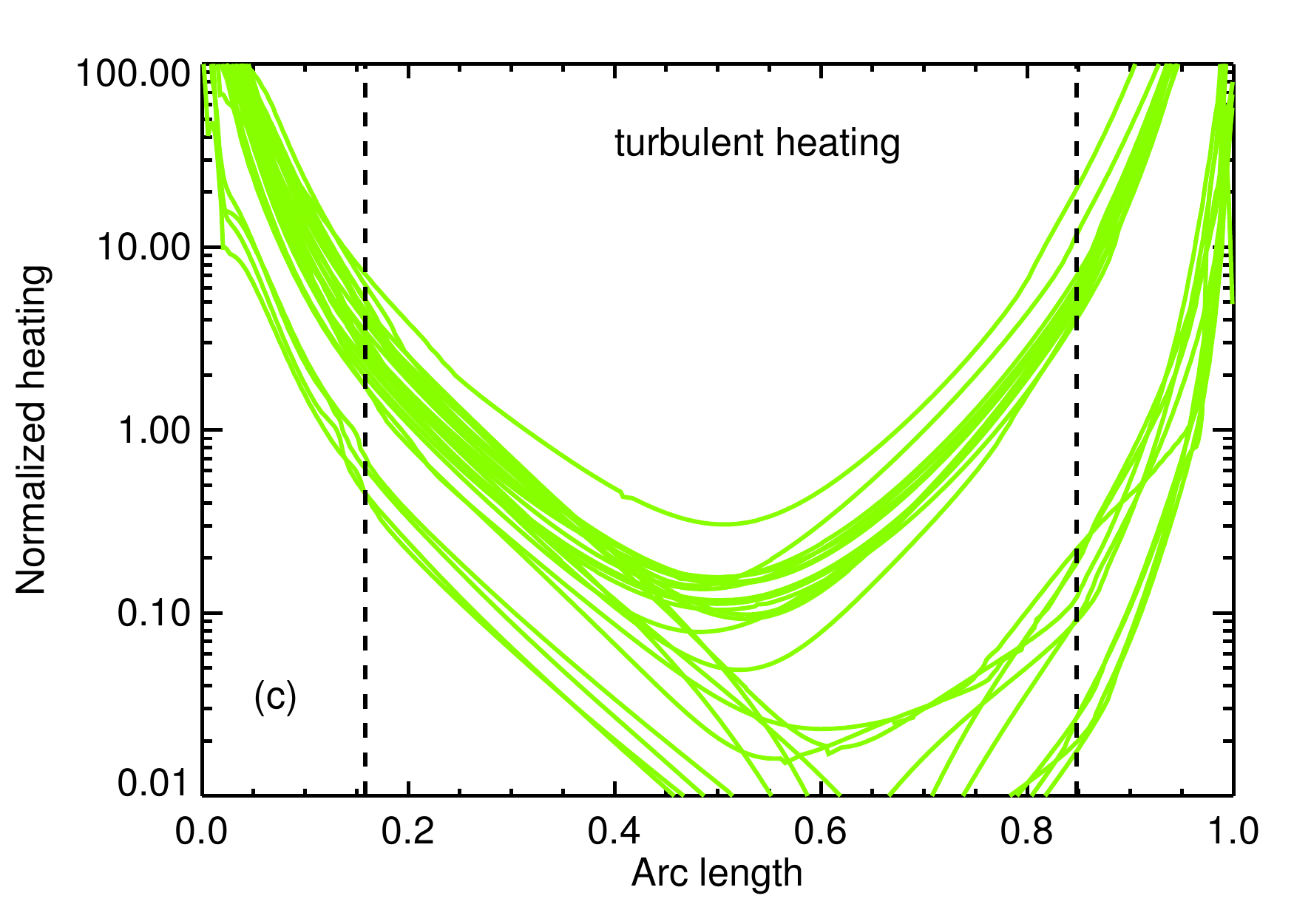}}
\caption{Similar to Fig. \ref{fig:set1_heating} but for the volumetric heating
rates of the ``random set'' of field lines marked in \fig{fig:loops-mid}
as green lines. Especially for the turbulent heating the asymmetry of the
field line shapes shows up clearly.
See \sect{S:random.set}.
\label{fig:set2_heating}}
\end{figure}

\begin{figure} 
\centerline{\includegraphics[width=0.8 \columnwidth]{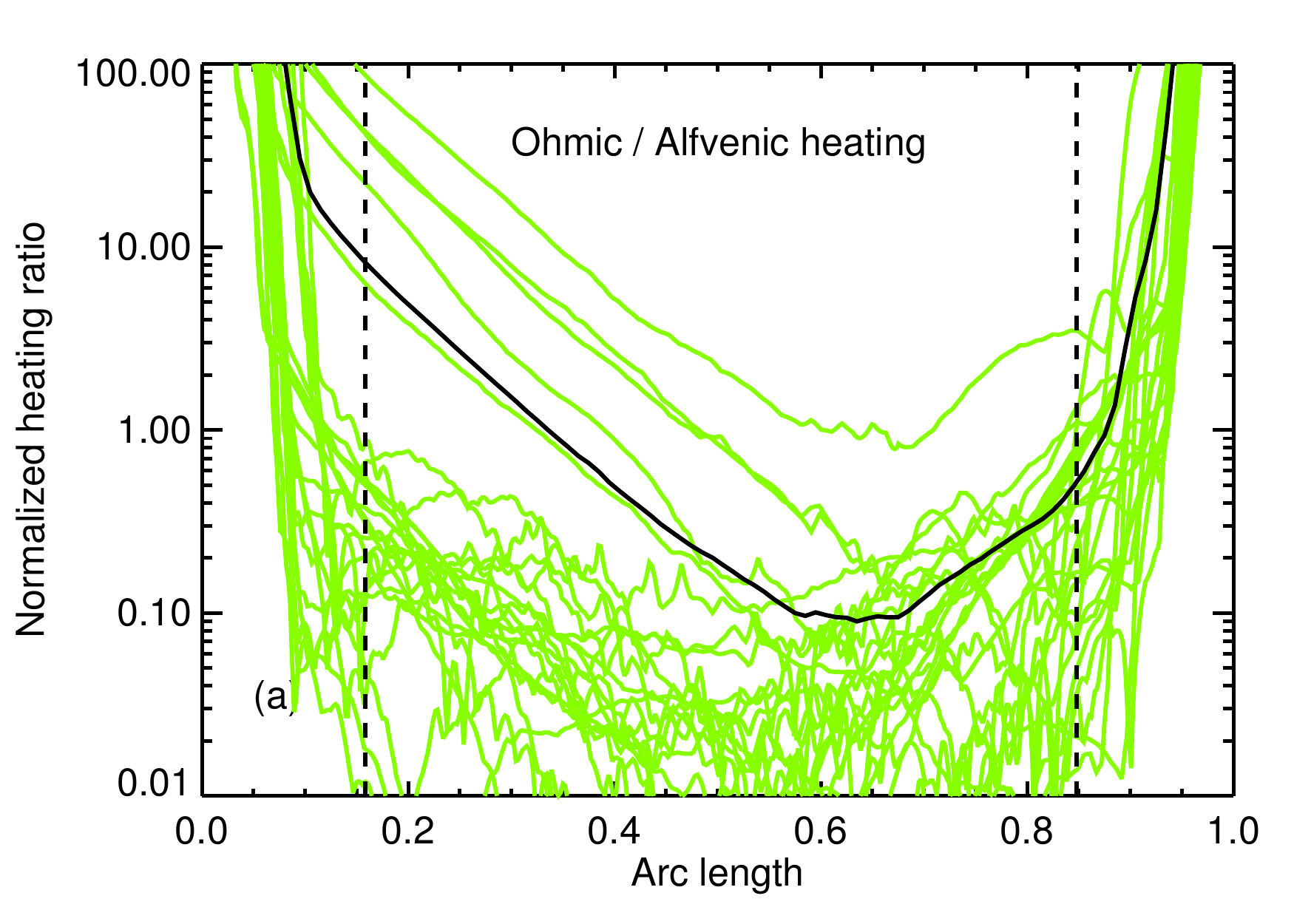}}
\centerline{\includegraphics[width=0.8 \columnwidth]{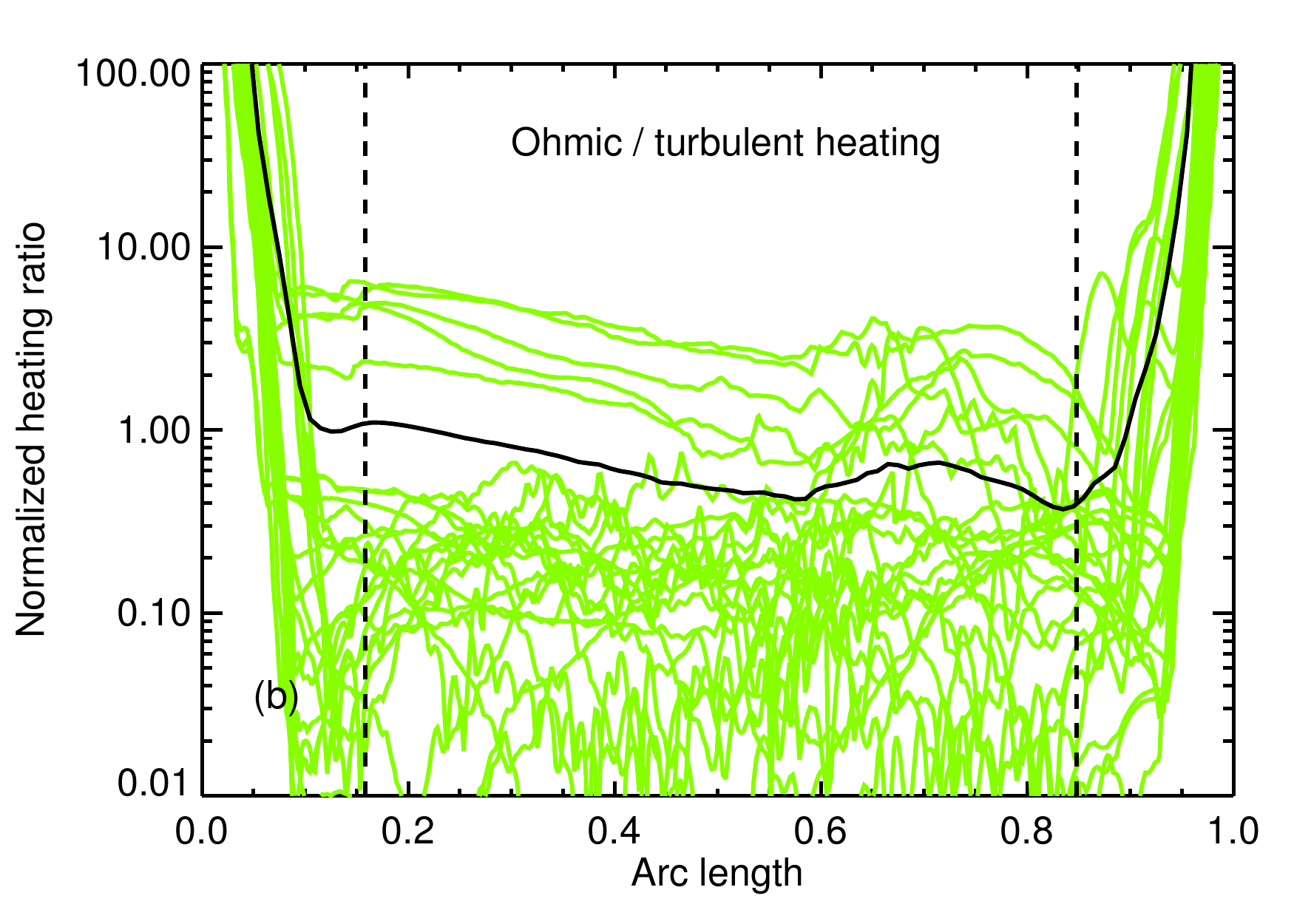}}
\caption{Similar to Fig. \ref{fig:set1_fractions} but for the ratios of the
heating rates of the ``random set'' of field lines marked in \fig{fig:loops-mid}
as green lines.
See \sect{S:random.set}.
\label{fig:set2_fractions}}
\end{figure}

The preceding discussion is for a quite special structure, namely for field lines associated with a bright loop. As a sort of blind test
we now investigate a more random set of field lines that are not associated to any particular coronal structures. This ``random set'' is plotted in green in Fig. \ref{fig:loops-mid} (see in \sect{S:field.lines} for the definition).
The volumetric heating rates and the ratio of the heating rates are plotted in \figs{fig:set2_heating} and \ref{fig:set2_fractions}

%
%

%

Interestingly, this selection gives overall similar results as the ``bright loop'' set, albeit with a larger scatter. This is a result of greater variety of field lines that sample different regions in the simulation box in different states. Despite the larger scatter, this clearly shows that the results outlined for the ``bright loop'' set can be generalized for the whole corona. This is not too surprising, because in a low-$\beta$ plasma the magnetic effects should be not too sensitive to the loading of the field lines with plasma. Thus field lines that are strongly loaded with hot plasma will show the same properties of the (magnetic) heating as other field lines that are not loaded with plasma. Of course, there is still the correlation between heating and coronal density that determines which field lines are loaded with how much plasma \citep[for a discussion of the appearance of loops see e.g.][]{Peter+Bingert:2012}.

There is one pointed difference between the ``bright loop''  and the ``random set'', though. The heating rates of the latter show stronger asymmetries between both sides of the loop,  which is particularly strong  for the turbulent heating parametrization (middle panel of \fig{fig:set2_heating}). This is due to the fact that some of the field lines in the ``random set'' are quite far from being semi-circular (cf. green lines in Fig. \ref{fig:loops-mid}). These more strangely-shaped field lines are hosting the asymmetric heating mainly because of the field strength asymmetry. Moreover, in the side where the field line does not reach very high into the corona, the density is higher and thus according to Eq.\,(\ref{eq:rappazzo}) the turbulent heating is stronger. This shows that the back-reaction of the heat input on the magnetic structure cannot be completely neglected (as noted in \sect{S:implementation}).

\subsection{From field lines to loop models}
We have looked at the spatial distribution of three different parametrizations of  coronal heating for two different selections of coronal field lines. The volumetric heating rates of all parametrizations drop for all field lines, and this drop is roughly exponential with height. The main difference is that the heat
input for the Ohmic and turbulent case is much more concentrated towards the foot points than in the case of Alfv\'enic heating.

It is instructive to explore whether this difference in foot point dominated and more  uniform heating has a significant effect on the coronal emission (and the dynamics). Therefore, we now synthesize coronal emission from 1D loop models with a spatial distribution of the heat input similar to the average of the set of field lines associated with the bright loop. Based on this we can investigate to what extent one can distinguish the heating parametrizations based on actually observable quantities. Obviously this can only be a first step, because in the end this has to be done in the framework of a 3D MHD model.

\section{One-dimensional coronal loop models}\label{S:one-d}

In the following we will construct simple 1D models of a coronal loop with constant cross-section and a prescribed heating function. All quantities depend only on the arc length along the magnetic field line defining the loop. The velocity is parallel to the loop. Besides the conservation of mass and momentum (including gravity), we solve the energy equation. The latter accounts for optically thin radiative losses \citep[following][]{Cook89} and heat conduction parallel to the magnetic field. This ensures that the coronal pressure is set self-consistently, which is pivotal if the resulting coronal emission radiated from the loop is to be synthesized, as we shall do here.
The 1D models are run using the Pencil Code \citep{brandenburg02} and follow the procedure of \cite{2012A&A...537A.152P}.

For the purpose of comparing the synthesized emission from the 1D loop we will adopt the average heating rate of the set of field lines associated with the bright loop (red lines in \fig{fig:loops-mid}, definition in
\sect{S:field.lines}, discussion in \sect{sect:hot_dense}).
This loop has a height of roughly 15\,Mm and a foot-point distance of about 28 Mm. This corresponds to  a roughly semi-circular shape with a length of about 45\,Mm which we will use in our numerical 1D model.

The volumetric heating rate $Q_i$ in the 1D model we  assume to fall off exponentially,
\begin{equation}
Q_i=H_{0,i}\,\exp\left(-\frac{z}{~\lambda_i~}\right),
\label{eq:heating}
\end{equation}
where $z$ denotes the geometric height and  $H_{0,i}$ is the heating rate at $z{=}0$. The scale height $\lambda_i$ for the heating remains to be determined for the three heating parametrizations, here represented by the index $i$.

\begin{figure} 
\centerline{\includegraphics[width=0.8 \columnwidth]{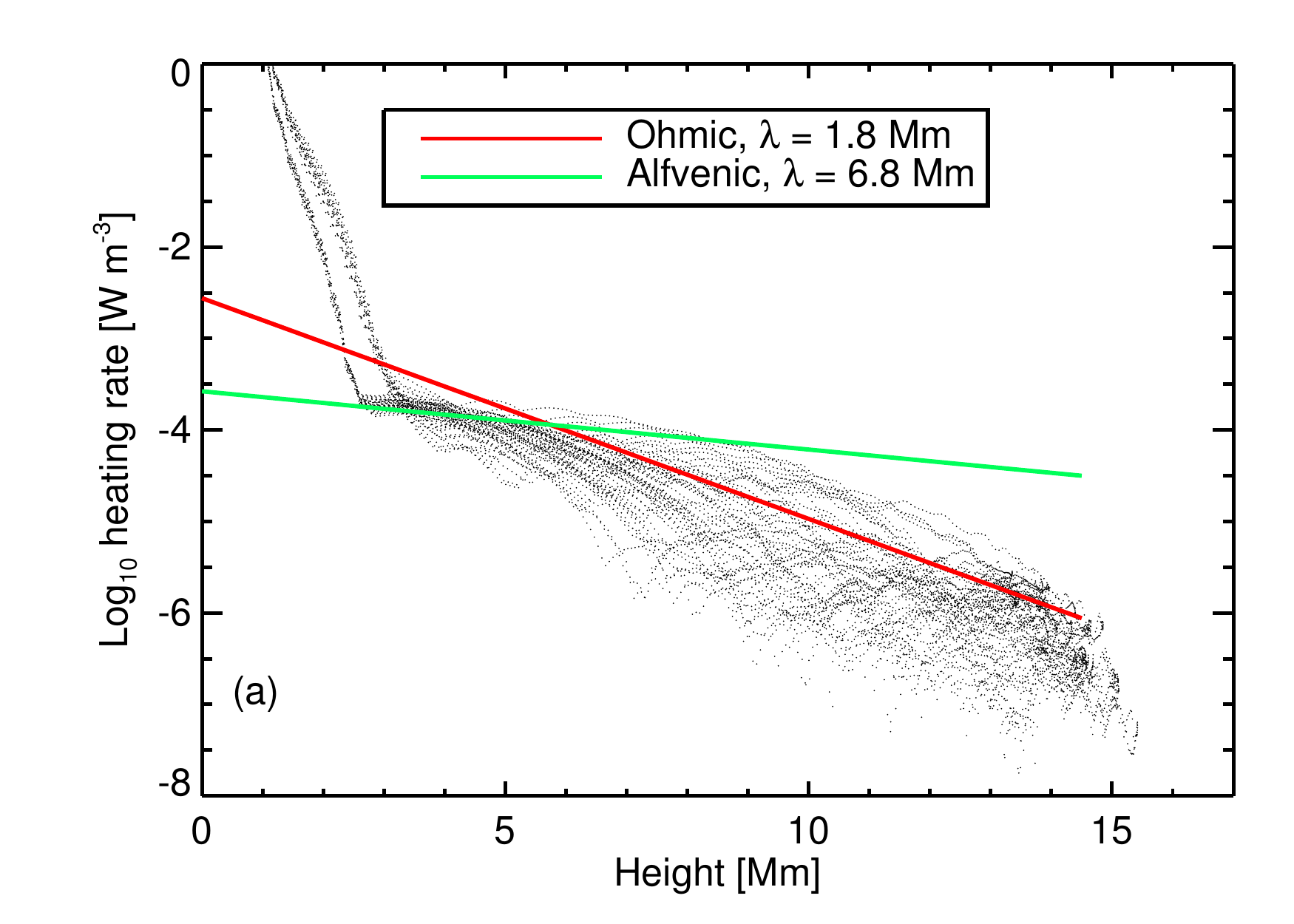}}
\centerline{\includegraphics[width=0.8 \columnwidth]{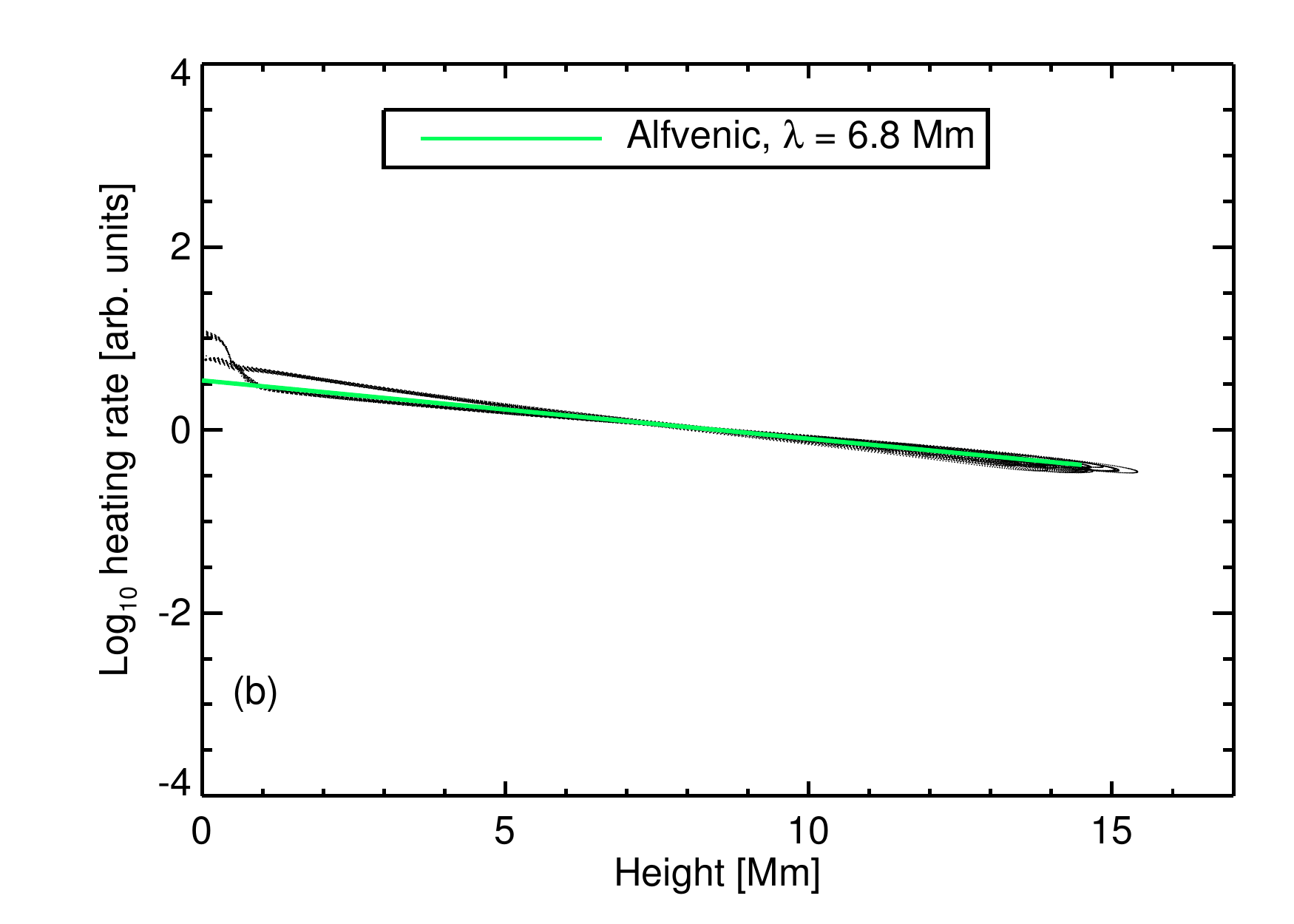}}
\centerline{\includegraphics[width=0.8 \columnwidth]{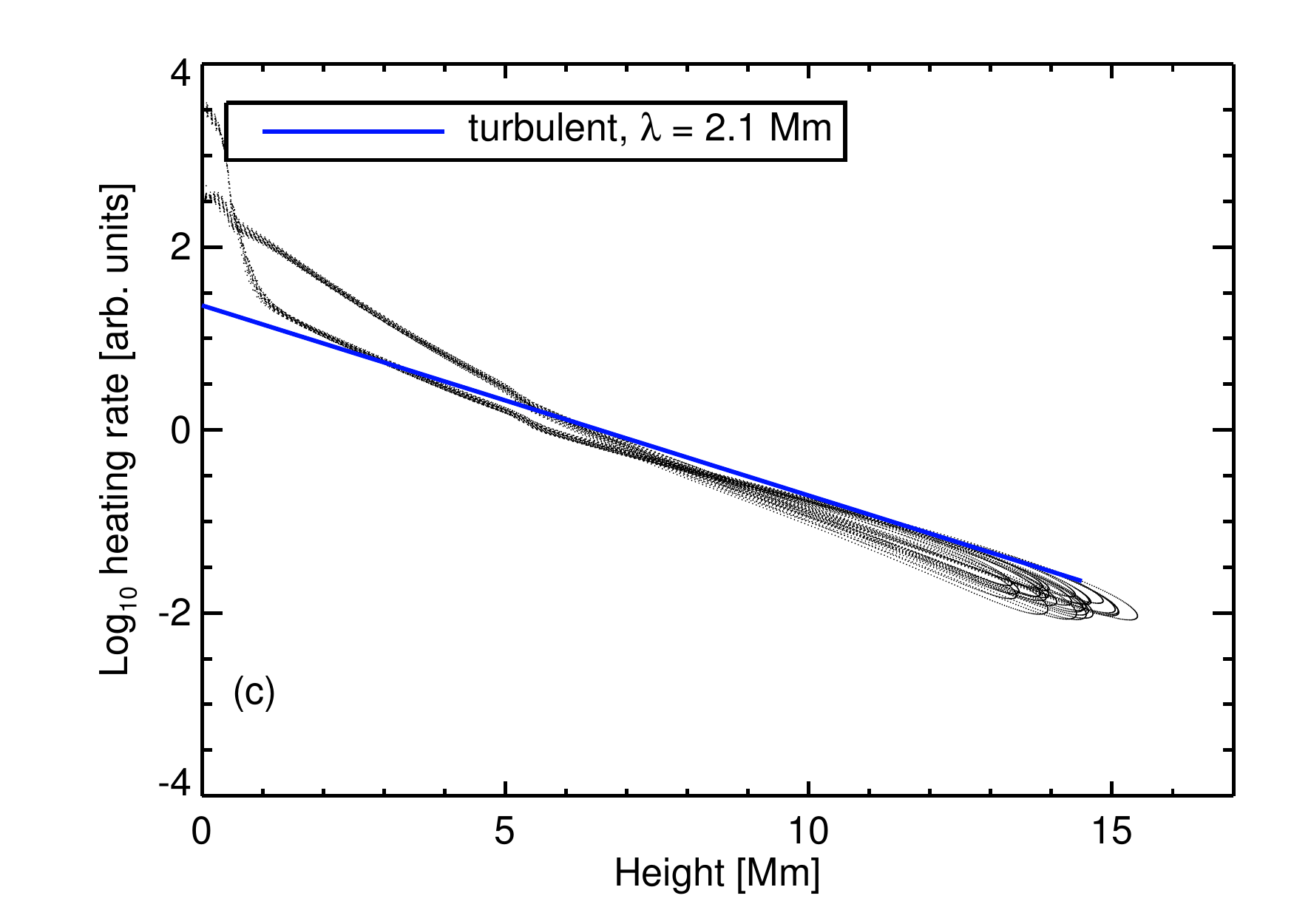}}
\caption{Volumetric heating rates over geometrical height for the ``bright loop'' (red in \fig{fig:loops-mid}) for the three parametrizations. The black dots show the actual heating rates along the field lines. The straight lines display the best exponential fit according to Eq.\,(\ref{eq:heating}). The resulting scale height for the heating rate, $\lambda_i$, is listed. For comparison the exponential drop of the Alfv\'enic heating is over-plotted (in arbitrary units).
See \sect{S:one-d}.
\label{fig:sqrfit1}}
\end{figure}

To determine the scale heights $\lambda_i$, we show in \fig{fig:sqrfit1}
the volumetric heating rates as a function of the geometric height $z$, including both loop legs, and not as a function of the arc length as before. We now fitted a simple exponential function in the form of Eq.\,(\ref{eq:heating}), which provides values for the $\lambda_i$.
For the fitting procedure we ignored all data points below the coronal base (at $\log{T}\,[\mathrm{K}]{=}5.5$, roughly at 3\,Mm).
These fits are over-plotted in \fig{fig:sqrfit1} and the values for the $\lambda_i$ are given. The exponential drop gives quite a good fit to the average variation. As suspected from the discussion in \sect{sect:hot_dense}, the Ohmic and turbulent heating show very similar results. This is why in the following we will only compare the Ohmic and Alfv\'enic parametrizations. In the Ohmic case we adopt a scale height of $\lambda_{\rm{Ohm}}{=}1.8$\,Mm, for the Alfv\'enic case $\lambda_{\rm{alf}}{=}6.8$\,Mm. For the Ohmic case we use a heating rate at $z{=}0$  of  about $H_{0,{\rm{Ohm}}}{\approx}3$\,mW/m$^{3}$ (see top panel of \fig{fig:sqrfit1}). The value for 
$H_{0,{\rm{alf}}}$ we determine using the requirement that the heat input into the corona (i.e.\ integrated above the coronal base) has to be the same in both cases. A more detailed description of the results of the 1D models are given in Appendix\,\ref{App:AppendixA}.
 
%
%

Because of the short scale height $\lambda_{\rm Ohm}$, the loop for the parametrization of the Ohmic heating is subject to a loss of equilibrium near the apex.  This process is well documented in the literature \citep[e.g.][and references therein]{Mueller+al:2003,Karpen+al:2006,2012A&A...537A.152P}. This leads to the episodic formation of condensations in the loop that eventually slide down into the photosphere. For the following discussion we thus investigate a snapshot in the comparably long time between two condensations (near $t{\approx}7500$\,s, see Appendix\,\ref{App:AppendixA}). The loop model of the Alfv\'enic heating with the longer scale height $\lambda_{\rm{alf}}$ reaches a static solution, which we then select for further analysis. For the times we analyse the two 1D models, in \emph{both} cases the velocities along the loop are very close to zero, less than 3\% of the sound speed.

%
\begin{figure*}
\sidecaption
\includegraphics[width=12cm]{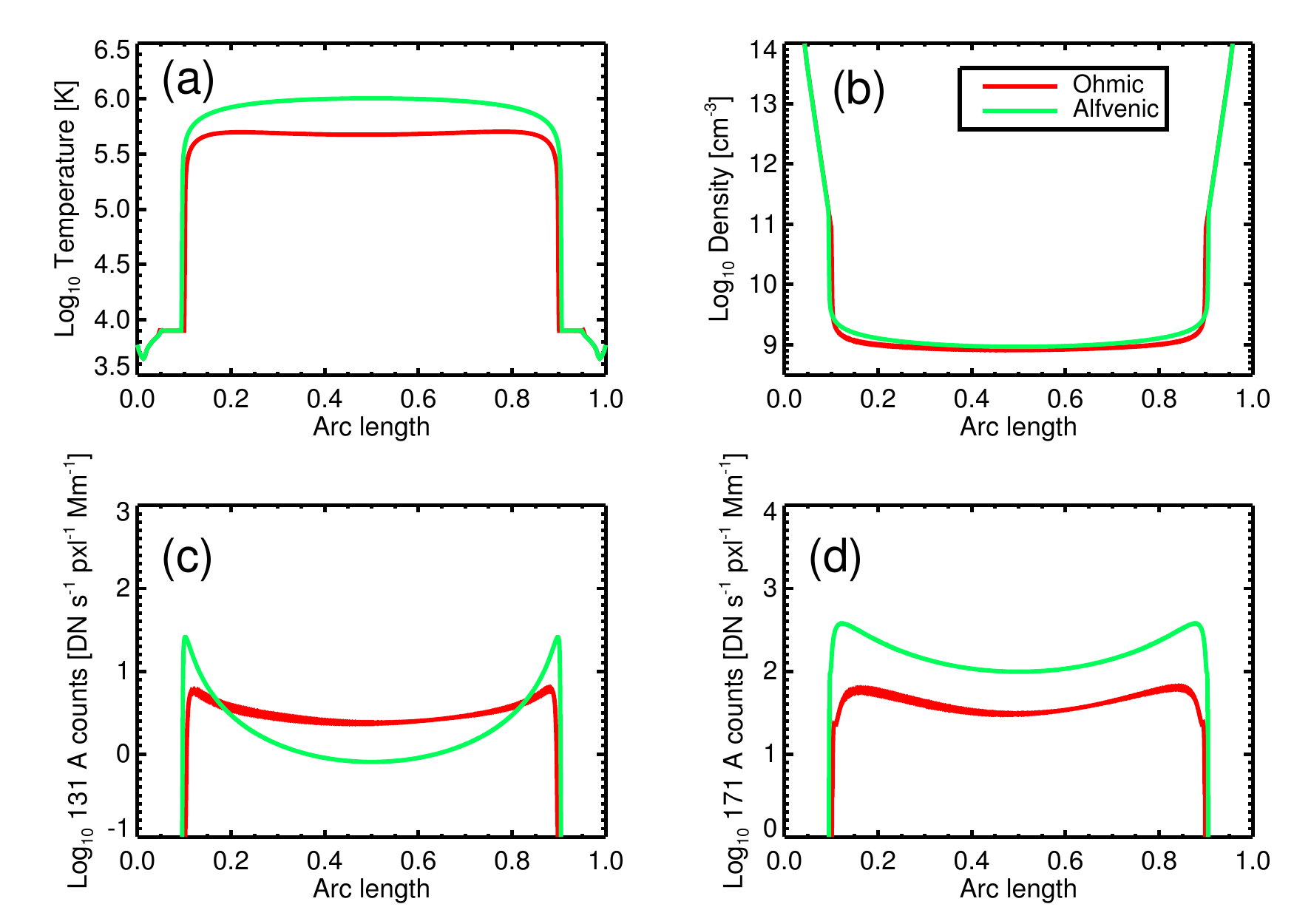}
\caption{Variation along the loops in the 1D models for the parametrization of the Ohmic (red) and Alfv\'enic heating (green). The arc length is normalized to the loop length of 45\,Mm.  The top panels show the temperature and the density, the velocities are very close to zero. The bottom panels show the coronal emission synthesized from the model as it would be observed with AIA in the 131\,\AA\ band (c) and the 171\,\AA\ band (d).
See \sect{S:one-d}.
\label{fig:loops_all}}
\end{figure*}
%

The temperature and density along the 1D model loops are plotted in \fig{fig:loops_all} (top row).
In the case of the Ohmic parametrization (in red),
the temperature is  below the temperature at the apex as found in the 3D model. The condensation that formed earlier on is an effective sink for the energy and thus the prevention of the strong condensation, e.g., by a siphon flow induced by asymmetric heating (see Appendix\,\ref{App:AppendixA}), could allow the temperature to reach higher values near the apex. However, the interest of the following discussion is not to what extent the 1D models can reproduce the results of the 3D model, but if one can find observable differences for the two loop models with different heating parametrizations.

For the  Ohmic and the Alfv\'enic heating we now calculate the coronal emission from the 1D loop model as it would be observed by the extreme UV imager on-board the Solar Dynamics Observatory, the Atmospheric Imaging Assembly \citep[AIA,][]{Lemen+al:2012}. In particular, following  the procedure of \citep{2012A&A...537A.152P} we synthesize the emission for the 131\,\AA\ and 171\,\AA\ channels that are dominated by emission from \ion{Fe}{viii} and \ion{Fe}{ix} from plasma  at temperatures of about 5.7 and 5.9 in $\log{T}$\,[K] \citep{Boerner+al:2012}. The resulting emission in these two pass bands for the two loop models is shown in the bottom panel of \fig{fig:loops_all}. 

Overall, the (relative) spatial distribution of the 171\,\AA\ emission is the quite similar for both cases, the Ohmic and the Alfv\'enic heating (\fig{fig:loops_all}\,c,d). However, because of a lower density (and temperature) the absolute level of the emission is different. Still, when  investigating actual observations to test which heating mechanism might be dominant, we would have to rely mostly on the relative distribution of the emission along the loop and not so much on the absolute level. Thus at least in the case we look at here, it would be hard to distinguish the heating mechanisms based on the 171\,\AA\ band alone.

 A clearer difference is seen in the 131\,\AA\ channel, where Alfv\'enic heating has two very clear ``horns'' near  the foot points of the loop, but is significantly weaker in the centre of the loop, when compared to the case of Ohmic heating. This is somewhat unexpected because the Ohmic heating shows a much stronger concentration towards the footpoints. However, the emission we see does not directly reflect the spatial variation of the heat input, but is a convolution of temperature and density, both of which are set by the heat input. Because of the higher apex temperature for the Alfv\'enic heating case, the spatial regions where the (comparatively cool) 131\,\AA\ emission originates is narrower and shifted down
when compared to the Ohmic case.

Depending on the band pass of coronal emission, two quite different heating mechanisms might produce similar or very different spatial distribution of the emission along the loop.
In this example the 171\,\AA\ band is similar, the 131\,\AA\ band is different. For other 1D loop experiments with different total heat input one can expect to find similar results, even though other channels might then be similar or different. So after all, observations of the coronal emission should hold the potential to distinguish different spatial distributions of the heat input, if one is not focusing on a single emission line or extreme UV bandpass alone.


\section{Conclusions}
In this paper we investigated differences and similarities of three  mechanisms to heat the corona: Ohmic heating following braiding of magnetic field lines by photospheric motions, the dissipation of Alfv\'en waves, and MHD turbulence. For our study we used the results of a  self-consistent 3D MHD simulation \citep{Bingert+Peter:2011}. From this  we calculated the Ohmic heating rate as resulting from that model, and the heating rates that would be given by parametrizations for  Alfv\'enic heating \citep{balleg2011} and turbulent heating \cite{Rappazzo+al:2008}.

We find that all the horizontally \emph{averaged} heating rates roughly drops exponentially with height. This is true for the average, and also when investigating the heating rate along individual field lines. Along magnetic field lines that are associated with a bright coronal loop in the 3D model, as well as along basically all other field lines reaching into the corona, we find a drop of the volumetric heating rate that is roughly exponential with height.

The Ohmic and the turbulent heating show roughly the same spatial distribution. This is not really surprising because the reduced MHD model for the turbulent heating is based upon  \cite{Rappazzo+al:2008}, is similar in principle to the full 3D MHD model \citep{Bingert+Peter:2011} --- in both models the field lines are braided and the non-linearity of the MHD equations drives the formation of small scales. This induces currents at small scales that are dissipated.
 While the full MHD model properly includes the energy equation, the reduced MHD models allows a much higher resolution. It is reassuring that these two models using different approaches give roughly the same result on the spatial distribution of the heating.
In contrast, the Alfv\'enic heating \citep{balleg2011}
shows a significantly smaller degree of concentration  of the heating rate towards the foot points.

Using the spatial distribution of the heat input from the 3D models we ran 1D loop models to make a first estimate if one can distinguish the different mechanisms by the distribution of the coronal emission along the loop.
For this we synthesized the emission how it would be seen with AIA. Here we find that some bands (for our example at 171\,\AA) look quite similar, while others (here at 131\,\AA) show quite different variations along the loop. 

The good news is that the different heating mechanisms will
produce different observables (when considering enough bands). However, the
bad news is that probably fiddling around with 1D models might be not sufficient
because there are to many free parameters. Here we showed only results for one loop for two AIA bands, and the situation is quite different for other loops and/or other bands.
Accounting for the spatial complexity, new 3D models with a self-consistent treatment of the heat input
based on driving in the photosphere will help in pinpointing the observational
similarities and differences of the different heating mechanisms.

\begin{acknowledgements}
This work was supported by the International Max-Planck Research School (IMPRS) on Physical Processes in the Solar System and Beyond. We thank the referee for the constructive 
\end{acknowledgements}

%

%
\newpage

\appendix


\section{Details of 1D coronal models}
\label{App:AppendixA}

%
\begin{figure*}
\sidecaption
\includegraphics[width=12cm]{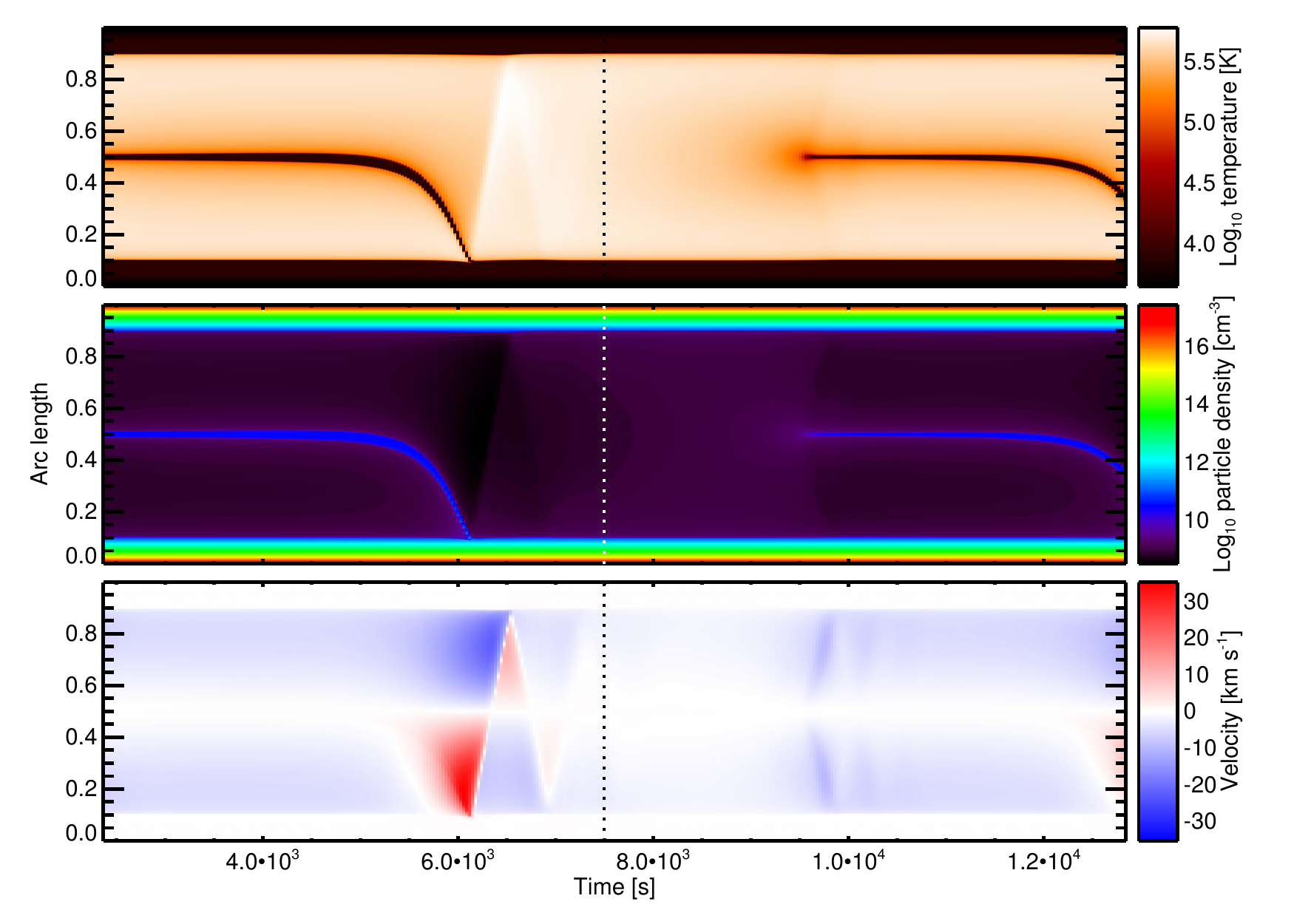}
\caption{Variation of temperature, density and velocity along the loop in the 1D model. Here we show the results for the parametrization of Ohmic heating with a scale height of $\lambda_{\rm{Ohm}}{=}1.8$\,Mm for an exponentially dropping heat input following Eq.\ref{eq:heating}. The arc length is normalized to the loop length of 45\,Mm. The vertical dashed line indicates the time of the snapshot that is further analysed in \sect{S:one-d}.
\label{fig:timeseries_ohm}}
\end{figure*}
%

\begin{figure*}
\sidecaption
\includegraphics[width=12cm]{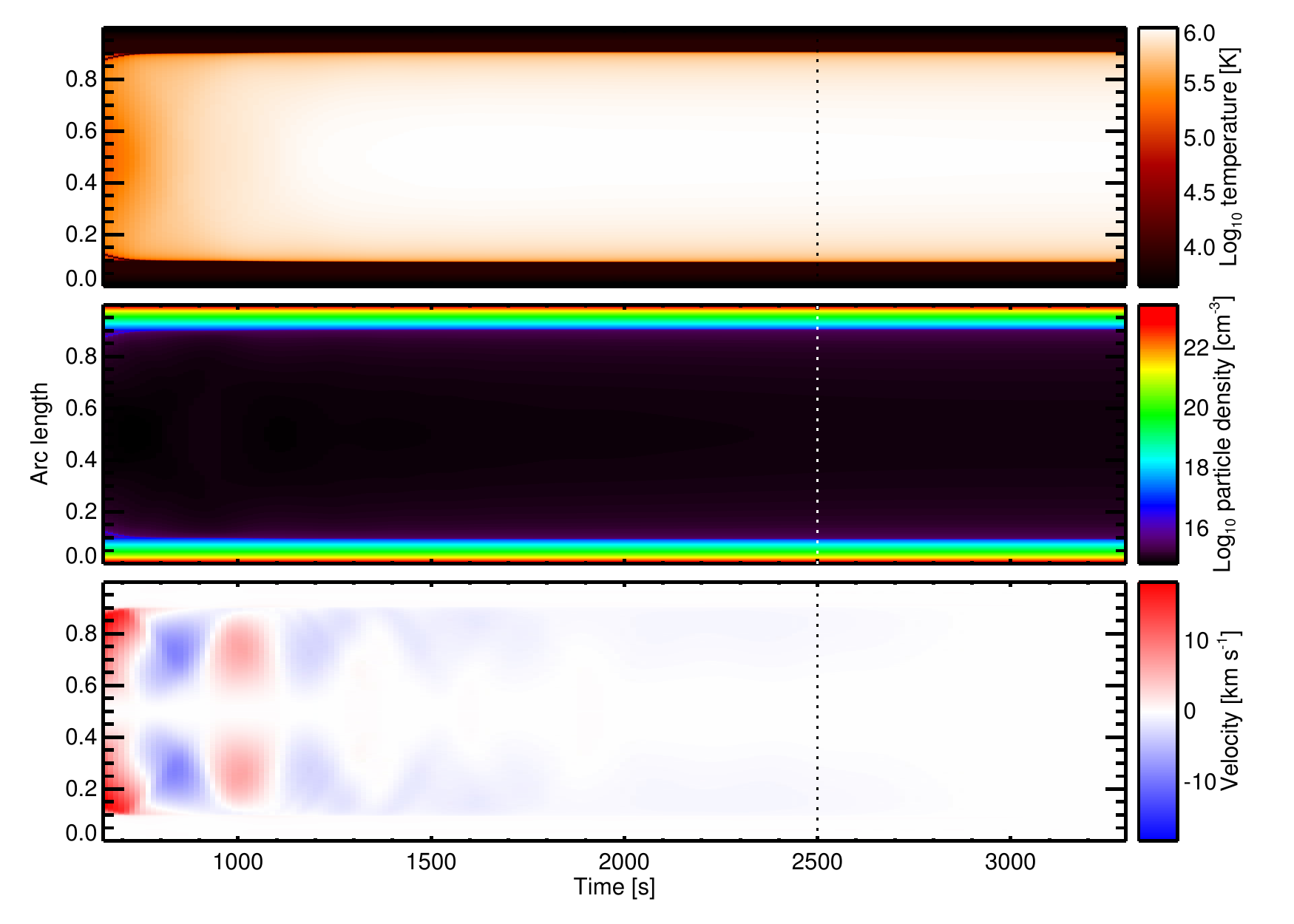}
\caption{Same as Fig. \ref{fig:timeseries_ohm}, but for the case of Alfv\'enic heating with a scale height of $\lambda_{\rm{alf}}{=}6.8$\,Mm.\label{fig:timeseries_alf}}
\end{figure*}
%

The 1D loop model describes a semi-circular 1D coronal loop with a length of 45\,Mm.
Along the loop we solve the mass and momentum balance including gravity. The energy balance includes heat
conduction parallel to the magnetic field and optically thin radiative losses. The heating is exponentially dropping according to Eq.\,(\ref{eq:heating})
with a scale height $\lambda_i$. Using the Pencil Code \citep{brandenburg02} we solve these equations on a 2048 grid, closely following \citep{2012A&A...537A.152P}. Starting from a initial condition with a prescribed temperature profile in hydrostatic equilibrium we evolve the equations.

The results of the numerical experiments are shown in \fig{fig:timeseries_ohm} for the parametrization of the Ohmic heating ($\lambda_{\rm{Ohm}}{=}1.8$\,Mm)
and in 
\fig{fig:timeseries_alf} for the Alfv\'enic heating  ($\lambda_{\rm{alf}}{=}6.8$\,Mm). Temperature, density, and velocity are plotted as a
space-time plot as a function of arc length along the loop (normalized to the loop length) and time.

In the case of Alfv\'enic heating (\fig{fig:timeseries_alf}) the coronal loop
reaches a stable static state after some 2000 s. For the discussion of this parametrization in \sect{S:one-d} we thus select a snapshot at $t{=}2500$\,s, where the velocity is very close to zero, less than 3\% of the sound speed.

The situation is different for the Ohmic heating (\fig{fig:timeseries_ohm}). The strong concentration of the heating towards the footpoints
evaporates chromospheric material that is then climbing up the loop towards the apex, where
the heating rate is quite low. As the density increases, the radiative losses
of the  plasma increase  and the plasma is effectively cooling. In a runaway process  then  a condensation forms because radiation becomes more efficient with decreasing temperature. Finally the condensation slides down one
side of the loop. Because the heat input is kept constant, the whole procedure starts again. This process of loss of equilibrium is well known  \citep[e.g.][]{Mueller+al:2003,Karpen+al:2006,2012A&A...537A.152P} and the results we find for the onset of the condensation (depending on $\lambda_i$) is consistent with previous studies. From inspection of \fig{fig:timeseries_ohm} it is clear that between condensations there are long stretches of time where the loop is comparably stable at hot temperatures without any condensation present. During the time span shown in \fig{fig:timeseries_ohm} this covers about 3000\,s. This is long compared to the sound-crossing time, which is of the order of 400\,s.
Arguing that we want to catch the loop undergoing Ohmic heating in a phase when it is hot and free of condensations we select a snapshot at $t{=}7500$\,s (cf.\ \fig{fig:timeseries_ohm}) for the further analysis in \sect{S:one-d}. At this time the velocity along the loop is almost zero everywhere, so that we consider this a quasi-steady state.

The 3D MHD model our study is based upon \citep{Bingert+Peter:2011} does not find such condensations in their synthesized emission. In that 3D model the heat input along the loop is \emph{not} symmetric, which is evident from the top panel of \fig{fig:set1_heating}. Even comparably small deviations from a symmetric heating will lead to different pressures at the coronal base of the two loop legs, which will drive a siphon flow through the loop \cite[e.g.][]{Boris+Mariska:1982}. To some extent, such a siphon flow will prevent strong condensations to form because the flow carries away the plasma as soon as it starts to condense (Z.\,Miki\'c, priv.\ comm.). Thus we do not see a condensation in our 3D model. Furthermore, because the condensations do not form, the loop apex can reach higher temperatures, as we see in the 3D model.

In the more idealized 1D loop model with a perfectly symmetric heating rate we see the condensations form. However, in the view of the above discussion it is reasonable to concentrate in on the phase between the condensations and use the snapshot at $t{=}7500$\,s for the analysis in \sect{S:one-d}.

\end{document}